\documentclass[aps,pra,twocolumn,amsmath,eqsecnum,amssymb,superscriptaddress,showpacs]{revtex4-1}

\usepackage{graphicx}
\usepackage{color}
\usepackage{epsfig}
\usepackage{amssymb}

\begin{document}

\title{Quantum entanglement in the one-dimensional spin-orbital SU(2)$\otimes XXZ$ model}

\author {    Wen-Long You }
\affiliation{Max-Planck-Institut f\"ur Festk\"orperforschung,
             Heisenbergstrasse 1, D-70569 Stuttgart, Germany }
\affiliation{College of Physics, Optoelectronics and Energy, Soochow University,
             Suzhou, Jiangsu 215006, People's Republic of China }

\author {    Peter Horsch }
\affiliation{Max-Planck-Institut f\"ur Festk\"orperforschung,
             Heisenbergstrasse 1, D-70569 Stuttgart, Germany }

\author {    Andrzej M. Ole\'s }
\affiliation{Max-Planck-Institut f\"ur Festk\"orperforschung,
             Heisenbergstrasse 1, D-70569 Stuttgart, Germany }
\affiliation{Marian Smoluchowski Institute of Physics, Jagiellonian University,
             prof. S. \L{}ojasiewicza 11, PL-30348 Krak\'ow, Poland }

\date{\today}

\begin{abstract}
We investigate the phase diagram and the spin-orbital entanglement of a
one-dimensional SU(2)$\otimes XXZ$ model with SU(2) spin exchange and 
anisotropic $XXZ$ orbital exchange interactions and negative exchange 
coupling. As a unique feature, the spin-orbital entanglement entropy in 
the entangled ground states increases here linearly with system size.
In the case of Ising orbital interactions we identify an emergent 
phase with long-range spin-singlet dimer correlations triggered by 
a quadrupling of correlations in the orbital sector. The peculiar 
translational invariant spin-singlet dimer phase has finite von Neumann 
entanglement entropy and survives when orbital quantum fluctuations are 
included. It even persists in the isotropic SU(2)$\otimes$SU(2) limit.
Surprisingly, for finite transverse orbital coupling the long-range 
spin singlet correlations also coexist in the antiferromagnetic spin 
and alternating orbital phase making this phase also unconventional.
Moreover we also find a complementary orbital singlet phase that 
exists in the isotropic case but does not extend to the Ising limit.
The nature of  entanglement appears essentially different from that
found in the frequently discussed model with positive coupling.
Furthermore we investigate the collective spin and orbital wave
excitations of the disentangled ferromagnetic-spin/ferro-orbital
ground state and explore the continuum of spin-orbital excitations.
Interestingly one finds among the latter excitations two modes of
exciton bound states. Their spin-orbital correlations differ
from the remaining continuum states and exhibit logarithmic
scaling of the von Neumann entropy with increasing system size.
We demonstrate that spin-orbital excitons can be experimentally
explored using resonant inelastic x-ray scattering, where the
strongly entangled exciton states can be easily distinguished 
from the spin-orbital continuum.
\end{abstract}

\maketitle

\section{Introduction}
\label{sec:int}

Spin-orbital coupling phenomena are ubiquitous in solids and have been
known to exist since the early days of quantum mechanics and band theory,
but only recently it was realized that the quantum nature of orbital
degrees of freedom plays a crucial role in the fields of strongly
correlated electrons \cite{Kug82,Fei97,Tokura,Ole05,Kha05,Ole12,Li14}
and cold atoms \cite{Zhao,Wu08,Sun12,Victor,Zho15}. The growing evidence 
of spin-orbital entanglement (SOE) accumulated due to novel experimental 
techniques which probe a variety of underlying electronic states. The 
strong Coulomb interactions and the relativistic spin-orbit interaction 
entangle locally the spin and orbital degrees of freedom \cite{Jac09} 
which display an amazing variety of fundamentally new and fascinating
phenomena, ranging from topologically nontrivial states \cite{Hasan},
relativistic Mott-insulating behavior in $5d$ \cite{Kim,Comin} and $4d$ 
\cite{Kha13,Akb14} transition-metal oxides and entanglement on
superexchange bonds in spin-orbital models \cite{Ole12,Ole06}. Other
more recent developments include entangled spin-orbital excitations
\cite{Brz14,Kag14}, doped spin-orbital systems \cite{Brz15}, skyrmion
lattices in the chiral metal MnSi \cite{Muh09}, multiferroics,
spin-Hall effects \cite{Xue13}, Majorana and Weyl fermions \cite{Wan11},
topological surface states \cite{Zhu13}, Kondo systems \cite{Gas13},
exotic spin textures in disordered systems, to name just a few.

To date, experimental observation of a dynamic spin-orbital state has
been a challenge. Apart from the intrinsic anisotropy and the relative 
complexity of the orbital couplings, it has been shown that the 
interplay between the two frustrated degrees of freedom may lead to 
exotic states of matter. An x-ray scattering study of a dynamic 
spin-orbital state in the frustrated magnet Ba$_3$CuSb$_2$O$_9$ 
supports spin liquid state \cite{Yuki,Kat15}, while FeSc$_2$S$_4$ 
\cite{Leon,Mit14,Lau14} and the $d^1$ effective models on the triangular 
lattice \cite{Nor08} and on the honeycomb lattice \cite{Karlo,Mil14} are 
found to be candidates for spin-orbital liquids in the theory.
Recently remarkable progress was achieved due to rapidly developed 
resonant inelastic x-ray scattering (RIXS) techniques \cite{vdB11} 
which helped to explore the elementary excitations in Sr$_2$CuO$_3$ 
\cite{Woh11,Sch12} and Sr$_2$IrO$_4$ \cite{Jungho}, with
antiferromagnetic (AF) and ferro-orbital (FO) order in ground states.
Orbital order in the spin-gapped dimerised system Sr$_3$Cr$_2$O$_8$
below the Jahn-Teller transition was also identified \cite{Zhe11}.
However, it remains challenging experimentally and theoretically, 
mainly owing to the lack of an ultimate understanding of spin-orbital 
correlations.

In the Mott insulators with an idealized perovskite structure, the
low-energy physics is described by spin-orbital models, similar to the
Kugel-Khomskii model \cite{Fei97}, where the spin and orbital are
considered on equal footing as dynamic quantum variables \cite{Ole05}.
Spin interaction possesses SU(2) symmetry, which will be broken however
by the relativistic spin-orbit coupling. It couples spins to the
orbitals, that are in general non-SU(2)-symmetric in a solid.
However, this coupling can frequently be neglected in realistic $3d$ 
systems and one is left in general with entangled spin-orbital 
superexchange problem \cite{Ole12}, that is the eigenstates cannot be 
written as products of spin and orbital wave functions. One immediate 
consequence of entanglement is that spin and orbital terms cannot be 
factorized in the mean-field approach. Orbitals are spatially 
anisotropic and thus their interactions have lower symmetry than the
spin ones which reflects the directional dependence of the orbital wave
functions. For the fixed occupation of orbitals, the magnitude and sign
of the spin-orbital superexchange interactions follow the classical
Goodenough-Kanamori rules \cite{Goode}, but quantum fluctuations change
them and make it necessary to consider spin-orbital interplay in
entangled states on exchange bonds \cite{Ole06}. Therefore, it is 
important to measure whether eigenstates are entangled or not.

A natural measure of SOE is the von Neumann entropy (vNE) which we
write first for the nondegenerate ground state $|\Psi_0\rangle$,
\begin{equation}
{\cal S}_{\rm vN}^0\equiv
-\textrm{Tr}_A\{\rho_A^{(0)}\log_2\rho_A^{(0)}\}.
\label{svn}
\end{equation}
Here we consider a system $\Omega$ composed of two non-overlapping
subsystems \cite{Amico}, i.e., $\Omega=A\cup B$, $A\cap B=\emptyset$,
and $\rho^{(0)}_A$ is the reduced density matrix. It is
obtained by integrating the density matrix over subsystem $B$, i.e.,
$\rho^{(0)}_A=\textrm{Tr}_B|\Psi_0\rangle\langle\Psi_0|$.
However, one has to realize that information contained in entanglement 
entropy depends crucially on how one partitions the Hilbert space of 
the system. To investigate SOE we use here as two subsystems $A$ and 
$B$ the spin and orbital degrees of freedom in the entire chain. 
Standard spin-orbital phases may have entanglement in only one sector 
and here we concentrate on joint SOE \cite{You12}. This choice is 
distinct from the one conventionally made when the system is separated 
into two spatially complementary parts \cite{Eisert}, for instance in 
frustrated spin chains \cite{Mas09} or in the periodic 1D Anderson 
model \cite{Hag15}.

Though much attention was devoted to the ground state in the past
\cite{Chen07}, it has been noticed only recently that the entanglement 
entropy of low-energy excitations may provide even more valuable 
insights \cite{You12,Rex,Ber13} which are of crucial importance to 
understand the origin of quantum phase transitions in spin-orbital 
systems \cite{You15}. The well known area law of the bipartite 
entanglement entropy restricts the Hilbert space accessible to a 
ground state of gapped systems \cite{Got10,Cal04}, while the area law 
is violated by a leading logarithmic correction in critical systems, 
whose prefactor is determined by the number of chiral modes and 
precisely given by Widom conjecture \cite{Hel11}. In this respect, 
the application of the entanglement entropy in describing quantum 
criticality in many-body Hamiltonian merits a lot of studies
\cite{Wu04,Amico}.

On the other hand, the excited states have the mixture of logarithmic
and extensive entanglement entropy, and the logarithmic states are
expected to be negligible in number compared to all the others. The
entanglement in excited state is proven always larger than that of 
the ground state of a spin chain \cite{Alb09}. 
For a spin-orbital coupled system, the division of spin and orbital
operators retains the real-space symmetries, which is beneficial to
the calculation of mutual entanglement. In two-particle states, the
SOE is determined by the inter-component coherence length \cite{You12},
as though the state has sufficient decay of correlations \cite{Mas09}.

The aim of this paper is to use the entanglement entropy to
investigate the full phase diagram of the one-dimensional (1D)
anisotropic spin-orbital SU(2)$\otimes XXZ$ model. The main motivation
for considering the Ising asymmetry in the orbital sector comes from
the observation that spin-orbital entanglement is large when both
subsystems, i.e., spin and orbital sectors, reveal strong quantum
fluctuations. Thus the Ising anisotropy which is present in many
physical systems introduces additional control of orbital fluctuations
and thereby provides an important control parameter for SOE.
Here we focus on the model with negative
exchange interaction. This choice of the exchange coupling restricts
somewhat joint spin-orbital fluctuations being particularly large near
the SU(4) symmetric point in the 1D spin-orbital model with
\textit{positive} coupling constant \cite{Li99}, but opens novel
possibilities for entangled states, as we show below \cite{You15}. 
An interesting phase with entangled ground state, consisting of 
alternating spin singlets along the spin-orbital ring, is found for 
Ising orbital interactions when the dimerization in the spin channel 
induces the change from FO to alternating orbital (AO) correlations. 
Here we report the complete phase diagram of the anisotropic 
SU(2)$\otimes XXZ$ spin-orbital model, with two phases of similar 
nature which gain energy from singlet correlations leading to 
dimerization, either in spin or in orbital sector. These phases were 
overlooked before in the fully symmetric case, i.e., in the phase 
diagram of the isotropic SU(2)$\otimes$SU(2) model \cite{You12}.

We also analyze the nature of spin-orbital excited states,
particularly in the case of the disentangled ferromagnetic (FM) and
FO ground state, labeled as FM/FO order. We also analyze entanglement
entropy in the excited states for the FM/FO phase and show that
spin-orbital excitations form a continuum, supplemented by collective
bound states. The latter states are characterized by a logarithmic 
scaling behavior, and as we show could be detected by properly 
designed RIXS experiments \cite{Ame09,Bis15,Che15}.

The paper is organized as follows. The model is introduced in Sec.
\ref{sec:som}. In Sec. \ref{sec:ising} we present an analytic solution
for the ground state in the Ising limit of the orbital interactions.
A more general situation with anisotropic $XXZ$ orbital interaction is
analyzed in Sec. \ref{sec:xxz}, and the phase diagram for the isotropic
SU(2)$\otimes$SU(2) model is reported in Sec. \ref{sec:xxx}. This model
and the obtained SOE are different from the AF case, as shown in Sec.
\ref{sec:III}. Next we determine the elementary excitations in the
FM/FO phase in Sec. \ref{sec:ff} and show that they are entangled 
although the ground state is disentangled. The vNE spectral function
is presented in Sec. \ref{sec:ffen}, including the scaling behavior of
the bound states which is contrasted with that in the AF/AO ground
state. In Sec. \ref{sec:rixs} we explore the possibilities of
investigating entanglement in the present 1D spin-orbital model by
RIXS. The paper is concluded by a discussion and brief summary in Sec.
\ref{sec:summa}. Some additional technical insights which are 
accessible by an exact solution of the two-site model are presented 
in the Appendix.

\section{The 1D spin-orbital SU(2)$\otimes XXZ$}
\label{sec:som}

We consider the 1D spin-orbital Hamiltonian which couples $S=1/2$
spins and $T=1/2$ orbital (pseudospin) operators,
\begin{equation}
{\cal H} =-J\,\sum_{j} {\cal H}_j^S(x)\otimes{\cal H}_j^T(\Delta,y)\,,
\label{som}
\end{equation}
with SU(2) spin Heisenberg interaction ${\cal H}_j^S(x)$, orbital 
anisotropic $XXZ$ interaction ${\cal H}_j^T(\Delta,y)$,
\begin{eqnarray}
\!\!\!\!\!\!\!\!
{\cal H}_j^S(x)&=&\vec{S}_{j} \cdot \vec{S}_{j+1} + x,   \\
\!\!\!\!\!\!\!\!
{\cal H}_j^T(\Delta,y)&=&
\Delta
\left(T_j^xT_{j+1}^x+T_j^yT_{j+1}^y\right)+T_j^zT_{j+1}^z+y.
\end{eqnarray}
We take below $J=1$ as the energy unit. The model Eq. (\ref{som})
has the following parameters:
(i) $x$ and $y$ which determine the amplitudes of orbital and spin
ferro-exchange interactions, $-Jx$ and $-Jy$, respectively, and
(ii) $\Delta$ which interpolates between the Heisenberg ($\Delta=1$)
and Ising ($\Delta=0$) limit for 
orbital interactions. When $\Delta=1$, the spin
and orbital interactions are on equal footing and the symmetry of the
Hamiltonian (\ref{som}) is enhanced to SU(2)$\otimes$SU(2) --- this
model describes a generic competition between FM and AF spin, 
and between FO and AO bond correlations \cite{You12}.

We emphasize that the coupling constant $-J$ is \textit{negative}, so
at large $x>0$ and $y>0$ it gives a disentangled FM/FO ground state,
see below --- therefore the model may be called in short FM. This
choice of the exchange coupling restricts somewhat joint spin-orbital
fluctuations being large near the SU(4) symmetric point,
$(x,y)=(0.25,0.25)$, in the 1D spin-orbital model with 
\textit{positive}, i.e., AF coupling constant \cite{Li99}, but opens 
other interesting possibilities for entangled states, as we have shown 
recently \cite{You15}. Both total spin magnetization $S^z$ and orbital
polarization $T^z$ are conserved, and time reversal symmetry leads to
the total momentum either $k=0$ or $k=\pi$.

Before analyzing the spin-orbital model of Eq. (1) in more detail, let
us summarize briefly the properties of the well known AF model, with
positive coupling constant $J$. The SU(4) symmetric Hamiltonian found 
at $(x,y)=(0.25,0.25)$ is an integrable model which can be solved in
terms of the Bethe \textit{Ansatz} \cite{Li99,Sut75}. Away from the
SU(4) symmetric points this choice of the coupling constant favors the
phases with spin-orbital order depending on the actual values of $x$
and $y$, and the phase diagram obtained by numerical methods includes
in general phases with all types of coupled spin-orbital order, i.e.,
FM/FO, AF/FO, AF/AO, and FM/AO, as well as the gapless spin-orbital
liquid phase near the SU(4) point \cite{Ori00,Yam00}. In addition,
Schwinger boson analysis gives phases with spin-orbital valence-bond
correlations and also spin valence bond and orbital valence-bond
phases \cite{Li05}.
The latter two show a tendency towards dimerised spin or orbital
correlations which occur here in the proximity of the SU(4) point.
For some special choice of parameters the model can be solved exactly:
(i) when $\Delta=1$ and $x=y=3/4$, the exact ground state is doubly
degenerate with the spins and the orbitals forming singlets on
alternate bonds, while
(ii) when $\Delta=0$, $x=3/4$ and $y=1/2$, the non-Haldane spin-liquid
ground state can be analytically obtained \cite{Kol98,Kol01}, and
(iii) several integrable cases were presented for interactions with
special symmetries \cite{Martins,Mila}, or
(iv) with $XY$ orbital interactions ($\Delta=\infty$) \cite{Brz14}.

The form of Eq. (\ref{som}) is not the most general one but is
representative for real spin-orbital systems with anisotropic orbital
interactions. In real systems the orbital part contributes by additional
superexchange terms which are not coupled to SU(2) spin interaction
\cite{Ole05}. For instance, in the case of $t_{2g}$ orbital degrees of
freedom as in the perovskite titanates or vanadates, the interactions
along the $c$ cubic axis involve the doublet of two orbitals active
along it, i.e., the $yz$ and $zx$ orbitals \cite{Kha01}; a similar
situation is encountered in a tetragonal crystal field of a quasi-1D
Mott insulator \cite{Kug15}, or for $p_x$ and $p_y$ orbitals of a 1D
fermionic optical lattice \cite{Zhao,Wu08,Sun12}.

\textit{A priori}, due to the quartic spin-orbital joint term,
$\propto(\vec{S}_{j}\cdot\vec{S}_{j+1})
[\Delta(T_j^xT_{j+1}^x+T_j^yT_{j+1}^y)+T_j^zT_{j+1}^z]$
in the Hamiltonian Eq. (\ref{som}) the spin-orbital interactions are
entangled, and the spin and orbital operators cannot be separated from
each other in the correlation function, except for some ground or
excited states in which the SOE vanishes.
The spin-orbital bond correlations (\ref{Ctot})
\begin{equation}
C^\mathrm{tot}_1\equiv\left\langle(\vec{S}_{j}\cdot\vec{S}_{j+1})
[\Delta(T_j^xT_{j+1}^x+T_j^yT_{j+1}^y)+T_j^zT_{j+1}^z]\right\rangle,
\label{Ctot}
\end{equation}
are uniform in the considered system and $C^\mathrm{tot}_1$ does not 
depend on the site index $j$. We investigate below these composite 
quartic correlations and show that they could also be surprisingly 
large. As an additional criterion of setting up the phase diagram, 
we use below the fidelity susceptibility which elucidates the change 
rate of ground states in the parameter space \cite{You07}. It serves 
as an order parameter to characterize the phase diagram of the 
anisotropic ($\Delta<1$) spin-orbital model (\ref{som}). 
The fidelity susceptibility is defined as follows,
\begin{eqnarray}
\chi_{\textrm F}(\lambda)\equiv -2\,\lim_{\delta\lambda\rightarrow 0}
\frac{\ln{\cal F(\lambda,\delta\lambda)}}{(\delta\lambda)^2},
\label{fis}
\end{eqnarray}
where the fidelity
\begin{eqnarray}
{\cal F} (\lambda,\delta\lambda)=
|\langle\Psi_0(\lambda)|\Psi_0(\lambda+\delta\lambda)\rangle|,
\end{eqnarray}
is taken along a certain path in the parameter space in the vicinity
of the point $\lambda\equiv\lambda(\Delta,x,y)$.

\section{Ising orbital interactions ($\Delta=0$)}
\label{sec:ising}

In the Ising limit of orbital interactions ($\Delta=0$) the Hamiltonian 
(\ref{som}) simplifies and has SU(2)$\otimes\mathbb{Z}_2$
symmetry --- it is a prototype model for the directional orbital
interactions with quenched quantum fluctuations in $t_{2g}$ systems.
This may happen in real compounds in two ways:
(i) either only one of the two active orbitals is occupied by one
electron and contributes in the hopping processes along the 180$^\circ$
bonds \cite{Dag08} or 90$^\circ$ bonds \cite{Che09}, or
(ii) the orbital degrees of freedom are quenched in the presence
of strong crystal field.
In both these cases the orbital exchange (orbital-flip) processes are
blocked and orbital interaction are of a classical Ising-like form.
Such Ising interactions are frustrated when they emerge in higher
dimension, as in the well-studied orbital compass model
\cite{Dou05,Brz10,Tro10} and in Kitaev model \cite{Kit06},
see also a recent review on the compass model \cite{Nus15}. It is 
now intriguing to ask what happens to the SOE in this case. It may be
still triggered by spin fluctuations while the model with Ising spin
interactions (\ref{II-Ham}) is classical.

\begin{figure}[t!]
\includegraphics[width=8.2cm]{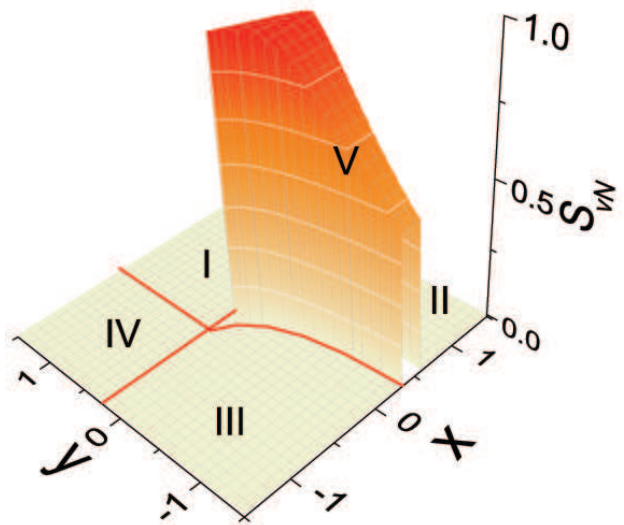}
\caption{(Color online)
Spin-orbital entanglement entropy ${\cal S}_\textrm{vN}^0$ Eq.
(\ref{svn}) and the phase diagram in the $(x,y)$ plane of the
SU(2)$\otimes\mathbb{Z}_2$ spin-orbital model (\ref{som}) with
$\Delta=0$ as obtained for the system size of $L=8$ sites. The critical
lines are discerned by both fidelity susceptibility and analytical
method. Phases I-IV are disentangled (${\cal S}_\textrm{vN}^0=0$) with
order 
defined as follows: FM/FO (phase I), AF/FO (phase II),
AF/AO (phase III), and FM/AO (phase IV).
The spin and orbital textures in phase V with finite entropy 
${\cal S}_\textrm{vN}^0>0$ are explained in the text. }
\label{fig:phd0}
\end{figure}

The phase diagram of the model Eq. (\ref{som}) at $\Delta=0$, i.e.,
in the absence of orbital fluctuations, which follows from fidelity
susceptibility (\ref{fis}) is displayed in Fig.~\ref{fig:phd0}.
As expected, one finds four trivial combinations of
spin-orbital order: FM/FO (phase I), AF/FO (phase II),
AF/AO (phase III), and FM/AO (phase IV). All these phases have the
entanglement entropy (\ref{svn}) ${\cal S}^{0}_\textrm{vN}=0$ and spins
and orbitals disentangle. Transitions between pairs of them are given
by straight lines and may be also obtained rigorously by the mean-field
approach. The ground state of a $L$-site chain stays in the subspace
$S^z=0$, $T^z=0$, momentum $k=0$ (always degenerate with $S^z=0$,
$T^z=0$, $k=\pi$ for all parameters) in phases III (AF/AO), IV (FM/AO)
and V, while it is found in the subspaces $S^z=0$, $T^z=\pm L/2$, $k=0$ 
in phases I (FM/FO) and II (AF/FO)
(of course, in phases I and IV also other values of $S^z\neq 0$,
with $-L/2\ge S^z\ge L/2$, are allowed and the ground states have the
respective degeneracy). The ground states with energy $E_0=0$ are
highly degenerate when $x<-1/4$ along the critical line $y=1/4$ between
phases III and IV, suggesting that antiparallel orbitals erase the spin
dynamics. Along the critical line $y=-1/4$ between phases I and II, the
ground states are also highly degenerate when $x\ge 3/4$, and parallel
orbitals on the bonds (in FO order) quench again the spin fluctuations.

Although the orbital interactions are Ising-like, entangled 
spin-orbital ground state occurs in phase V.
In order to understand better emergent phase V, we introduce the
longitudinal equal-time spin/orbital structure factor, defined for
a ring of length $L$ (with a lattice constant $a=1$; we use periodic
boundary conditions) by
\begin{eqnarray}
\label{Szzk}
S^{zz}(k)&=&\frac{1}{L}\sum_{j,j'=1}^{L}
e^{-ik(j-j')}\langle S_j^z S_{j'}^z\rangle, \\
\label{Tzzk}
T^{zz}(k)&=&\frac{1}{L}\sum_{j,j'=1}^{L}
e^{-ik(j-j')}\langle T_j^z T_{j'}^z\rangle.
\end{eqnarray}
The calculation of the equal-time structure factor $S^{zz}(k)$ for a
model of uncorrelated nearest neighbor dimers was compared with the one
for the kagome lattice ZnCu$_3$(OD)$_6$Cl$_2$ \cite{Han}. One finds
analytically that in the case $\Delta=0$, a cosine-like spin structure 
factor, i.e., $S^{zz}(k)\propto (1-\cos k)$, is revealed in phase V for 
$y=-1/4$, implying that only nearest neighbor spins are correlated. 
This finding is essential as the short-range spin correlation indicates 
here a translation invariant dimerised spin-singlet state which has the 
same spin structure as the Majumdar-Ghosh (MG) spin state \cite{Maj69}.
However, this state is not triggered here by frustrated interactions
$J_1$ and $J_2$, but is evidently induced by the correlations in the
orbital sector.

\begin{figure}[t!]
\includegraphics[width=7.2cm]{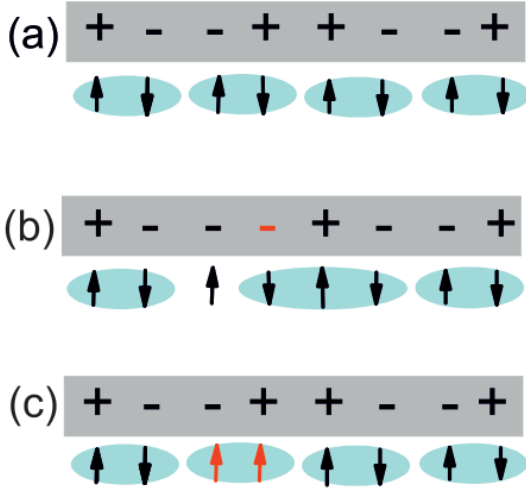}
\caption{(Color online)
(a) One of four translational equivalently spin and orbital
configurations in the Ising limit of the spin-orbital model
(\ref{som}) at $\Delta=0$ and $y=-0.25$. The spins form isolated
dimers (shaded ovals).
(b) A single orbital excitation and induced spin configuration.
(c) A single spin flip makes a singlet-triplet spin excitation,
but does not induce any change in orbital correlations. }
\label{fig:scheme}
\end{figure}

In the Ising limit we obtain the analytic ground state for phase V as
described below. The essential feature is that the energy is gained by
spin singlets occupying the bonds with AO states, while the bonds
connecting two spin singlets have FO order, see Fig. \ref{fig:scheme}(a).
To construct the ground state, we introduce the corresponding four
configurations in the orbital sector:
\begin{eqnarray}
\vert\phi_1\rangle&=&\vert + + - - + + - - \cdots \rangle, \nonumber \\
\vert\phi_2\rangle&=&\vert - + + - - + + - \cdots \rangle, \nonumber \\
\vert\phi_3\rangle&=&\vert - - + + - - + + \cdots \rangle, \nonumber \\
\vert\phi_4\rangle&=&\vert + - - + + - - + \cdots \rangle.
\end{eqnarray}
The solutions are classified by the momenta corresponding to the
translational symmetry of the system. The orbital wave functions in the
ground state for the momenta $k=0,\pi/2,3\pi/2,\pi$ correspond to:
\begin{equation}
\vert\phi_{k}\rangle=\frac{1}{2}\left(\vert\phi_1\rangle+e^{ik}
\vert\phi_2\rangle+e^{2ik}\vert\phi_3\rangle+e^{3ik}\vert\phi_4\rangle
\right).
\end{equation}
In spin subspace there are two distinct (but nonorthogonal) states:
\begin{eqnarray}
\vert\psi_1^{D}\rangle&=&[1,2][3,4]\cdots[N-1,N],        \nonumber \\
\vert\psi_2^{D}\rangle&=&[2,3][4,5]\cdots[N,1],
\end{eqnarray}
where the singlets are located on odd (even) bonds. Here a singlet is
defined by
$[l,l+1]=(\vert\uparrow\downarrow\rangle
         -\vert\downarrow\uparrow\rangle)/\sqrt{2}$.

One representative component of the ground state with the orbital part
$\vert\phi_4\rangle$ accompanied by the spin state $\vert\psi_1^D\rangle$
is shown in Fig. \ref{fig:scheme}(a).
The ground state in the $k=0$ subspace is given by the superposition
\begin{eqnarray}
&&|\Phi_{k=0}\rangle=                                     \nonumber \\
& &\!\frac{1}{2}\!\left(
 \vert\phi_1\rangle\!\otimes\!\vert\psi_2^{D}\rangle
\!+\!\vert\phi_2\rangle\!\otimes\!\vert\psi_1^{D}\rangle
\!+\!\vert\phi_3\rangle\!\otimes\!\vert\psi_2^{D}\rangle
\!+\!\vert\phi_4\rangle\!\otimes\!\vert\psi_1^{D}\rangle\right)  \nonumber \\
&=&\!\frac{1}{\sqrt{2}}\left(
 \frac{\vert\phi_1\rangle+\vert\phi_3\rangle}{\sqrt{2}}
 \otimes\vert\psi_2^{D}\rangle
+\frac{\vert\phi_2\rangle+\vert\phi_4\rangle}{\sqrt{2}}
 \otimes\vert\psi_1^{D}\rangle\right).
\end{eqnarray}
The state $|\Phi_{k=0}\rangle$ is entangled both in individual spin and
orbital subspaces, and also is characterized by SOE along the chain.
Such a many-body state, and similar states obtained for other momenta,
$k=\pm\pi/2$ and $k=\pi$, give an exact value of the vNE,
${\cal S}_{\rm vN}^{0}=1$. The resulting fluctuations between these
states suppress conventional order and the system features finite
entropy even at zero temperature, in contrast to the naive expectation
from the third law of thermodynamics. The emergent excitations are
also entangled and fundamentally different from the individual spin or
orbital ones, see Figs. \ref{fig:scheme}(b) and \ref{fig:scheme}(c).

\begin{figure}[t!]
\includegraphics[width=8.4cm]{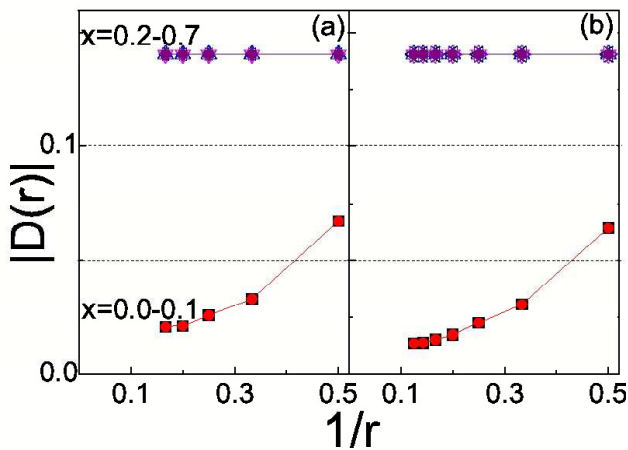}
\caption{Dimer correlation function $D(r)$ (\ref{Dr}) obtained
for the anisotropic SU(2)$\otimes\mathbb{Z}_2$ spin-orbital model  
for different values of $x\in[0,0.7]$ in phases V and III and
for the ring of length:
(a) $L=12$, and
(b) $L=16$ sites.
Parameters: $y=-0.25$ and $\Delta=0$. }
\label{fig:dimco}
\end{figure}

As the orbital correlations are classical in the Ising limit, we can
determine all the phase boundaries analytically by considering the spin
interactions for various orbital configurations. The lower boundary
between phase III (AF/AO) and V at $y=-1/4$ can be determined by 
comparing the uniform state with energies of AO correlation on a bond, 
i.e., $\langle T_j^zT_{j+1}^z\rangle=-1/4$, with the alternating state 
of pairs of the same orbitals shown in Fig. \ref{fig:scheme}(a), i.e.,
$\langle T_j^z T_{j+1}^z\rangle=(-1)^j/4$, which coexists with
spin dimer order (spin interactions vanish for a pair of identical
orbitals). One finds the following effective spin Hamiltonian in
this case:
\begin{eqnarray}
H_{{\rm DIM}}&=&\frac12\sum_{j\in{\rm odd}}
\left(\vec{S}_{j}\cdot\vec{S}_{j+1}+x\right),
\end{eqnarray}
and the corresponding ground state energy per site in the
thermodynamic limit is
\begin{eqnarray}
E^{0}_{{\rm DIM}}&=&\frac{1}{4}\left(-0.75+x\right).
\end{eqnarray}
The dimerised phase competes with the AO order coexisting with the
1D resonating valence-bond spin state, with energy
\begin{equation}
E^{0}_{{\rm AO}}=\frac{1}{2}\left(-0.4431+x\right). \\
\end{equation}
Hence, one finds that $E^{0}_{{\rm DIM}}<E^{0}_{{\rm AO}}$ for
$x>0.136$. The quadrupling due to spin-orbital interplay in phase V is
well seen by the calculation of the four-spin correlation function
which we define following Refs. \cite{Yu96,Ner11},
\begin{eqnarray}
D(r)&=&\frac{1}{L}\sum_i\left[\left\langle(\vec{S}_i\cdot\vec{S}_{i+1})
              (\vec{S}_{i+r}\cdot\vec{S}_{i+r+1})\right\rangle\right.
\nonumber \\
&-& \left.\left\langle\vec{S}_{i}\cdot\vec{S}_{i+1}\right\rangle
    \left\langle\vec{S}_{i+r}\cdot\vec{S}_{i+r+1}\right\rangle\right].
\label{Dr}
\end{eqnarray}
If $y=-1/4$, spin dimer correlations alternate and
\begin{equation}
D(r) = (-1)^r \left(\frac{3}{8}\right)^2,
\label{DV}
\end{equation}
which follows from Eq. (\ref{Dr}) for the alternating spin singlets,
$\langle\vec{S}_i\cdot\vec{S}_{i+r}\rangle=-3[1-(-1)^r]/8$.
Indeed, one finds this value (\ref{DV}) for $x\in[0.2,0.7]$ and the
result is robust and the same for systems sizes $L=12$ and $L=16$, see
Fig. \ref{fig:dimco}. On the contrary, for $x<0.2$ the values of $D(r)$
decrease with increasing distance $r$, and would vanish in the
thermodynamic limit of $L\to\infty$.

When $y<-1/4$, there are three competing phases with predetermined
orbital configurations (AO, DIM, or FO) and the corresponding spin
interactions given by effective spin Hamiltonians:
\begin{eqnarray}
\label{hao}
H_{\rm AO} &=&\left(\frac14-y\right)\sum_j(\vec{S}_j\cdot\vec{S}_{j+1}+x), \\
\label{hdim}
H_{\rm DIM}&=&\left(\frac14-y\right)\sum_{j\in{\rm odd}}
                (\vec{S}_{j} \cdot \vec{S}_{j+1} + x)                \nonumber \\
&-&\left(\frac14+y\right)\sum_{j\in{\rm even}}(\vec{S}_j\cdot\vec{S}_{j+1}+x), \\
\label{hfo}
H_{\rm FO}&=&-\left(\frac14+y\right)\sum_j(\vec{S}_j\cdot\vec{S}_{j+1}+x) .
\end{eqnarray}
In this case also even inter-singlet bonds contribute to the energy in
the $|{\rm DIM}\rangle$ state, but the spin correlations vanish, i.e.,
$\langle\vec{S}_j\cdot\vec{S}_{j+1}\rangle=0$.
It is obvious that $H_{\rm AO}$ and $H_{\rm FO}$ stand for the same
(translational invariant) spin Hamiltonian, and $H_{\rm FO}$ will have
lower ground state energy when
$x>-\langle\vec{S}_{j}\cdot\vec{S}_{j+1}\rangle_{\rm AF}\simeq 0.4431$.
The dimerised AF Heisenberg chain (\ref{hdim}) related to spin-Peierls
state cannot be solved trivially, with the exception of the free-dimer
limit ($y=-0.25$) and the uniform Heisenberg limit ($y=-\infty$)
\cite{Soos}, One finds the ground state energy per site
$\varepsilon_\infty(\delta)$ of a pure dimerised spin chain \cite{Spr86},
\begin{eqnarray}
\varepsilon_\infty(\delta)=
\frac{3}{4}\frac{1}{1+\alpha}\left(
1+\frac{\alpha^2}{8}+\frac{\alpha^3}{32}+\cdots\right),
\end{eqnarray}
with $\alpha=(1-\delta)/(1+\delta)$ and $\delta\equiv 1/\vert
4y\vert\gtrsim 0.4$. For $\delta\lesssim 0.4$,
\begin{eqnarray}
\varepsilon_\infty(\delta)=\ln2 -(\ln2-1)\vert\delta\vert^{4/3}.
\end{eqnarray}
In such a case, $E^{0}_{{\rm DIM}}=y[\varepsilon_\infty(\delta)-x]$.
The overwhelming dimerised phase will persist in a range of negative
values of $y$, and the boundaries close at $y=-\infty$, as is indicated
by structure factors and fidelity susceptibility.

The phase transitions in the phase diagram of Fig. \ref{fig:phd0} imply
the discontinuous changes of order parameters in first-order quantum
phase transitions. The orbital order changes from phase II (AF/FO) to 
phase III (AF/AO), as shown in Ref. \cite{You15}, but the N\'{e}el 
order persists in both of them and manifests itself in the two-spin 
correlation, $\langle S_i^zS_{i+r}^z\rangle$.
For translational invariant and orthonormal linear combinations of
the symmetry-broken N\'{e}el (AF) states,
\begin{eqnarray}
\vert\Phi_1^{\rm AF}\rangle&=&\vert
\uparrow\downarrow\uparrow\downarrow\cdots\uparrow\downarrow\rangle,
\nonumber \\
\vert\Phi_2^{\rm AF}\rangle&=&\vert
\downarrow\uparrow\downarrow\uparrow\cdots\downarrow\uparrow\rangle,
\label{afs}
\end{eqnarray}
there are spin $\langle S_i^zS_{i+r}^z \rangle= (-1)^r/4$ and
dimer $D(r)=0$ correlations (for $r\neq 0$), while for dimer states
$\vert \Phi_1^{\rm DIM}\rangle$ and
$\vert \Phi_2^{\rm DIM}\rangle$,
$\langle S_i^z S_{i+r}^z \rangle$=0 (for $r\neq\pm 1$) and $D(r)\neq 0$
(for $r\neq 0$), see Fig. \ref{fig:dimco}, respectively. These results
reflect the long-range nature of the two types of order. The AF
classical spin correlations (\ref{afs}) are replaced by a power law for 
the AF spin $S=1/2$ chain in the thermodynamic limit,
\begin{eqnarray}
\left\langle {\vec S}_i\cdot {\vec S}_{i+r}\right\rangle\sim
(-1)^{r}\frac{\sqrt{\ln\vert r\vert}}{\vert r\vert},
\end{eqnarray}
which is equivalently revealed by the structure factors $S^{zz}(k)$ and
$T^{zz}(k)$ defined by Eqs. (\ref{Szzk}) and (\ref{Tzzk}).

\section{Entanglement in the ground states}

\subsection{The anisotropic orbital interactions ($0<\Delta\le 1$)}
\label{sec:xxz}

When $\Delta=0$, there is no dynamics in the orbital sector, and the
orbital structure factor is dominated by a single mode which 
follows from 
the $\mathbb{Z}_2$ symmetry \cite{You15}. This changes when
$0<\Delta\le 1$ and the quantum fluctuations in the orbital sector
contribute. In order to understand the modifications of the phase
diagram in the entire interval $0<\Delta\le 1$, we select $\Delta=0.5$ 
and study the longitudinal equal-time spin/orbital structure factor, 
defined for a ring of length $L$ in Eqs. (\ref{Szzk}) and (\ref{Tzzk}).

\begin{figure}[b!]
\includegraphics[width=9.2cm]{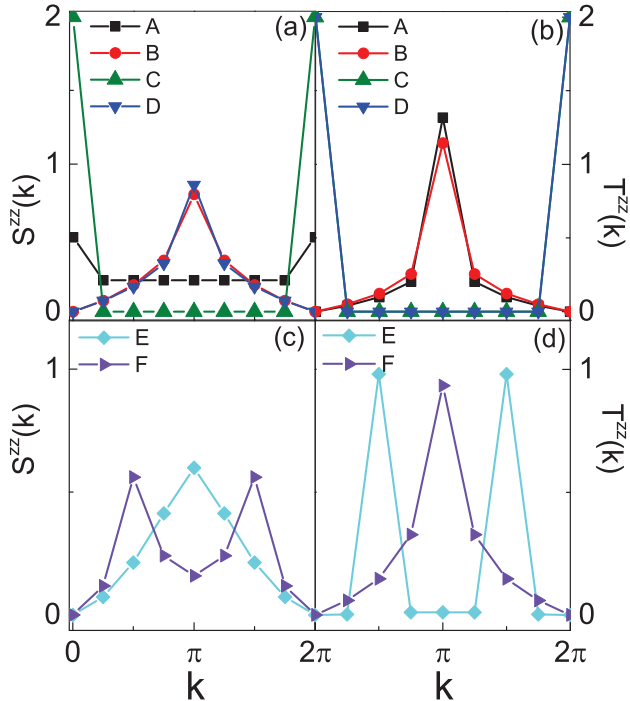}
\caption{(Color online)
The spin $S^{zz}(k)$ (a,c), and
orbital $T^{zz}(k)$ (b,d)
structure factors obtained for the selected points shown in the phase 
diagram of the anisotropic spin-orbital SU(2)$\otimes XXZ$ model, see 
Fig. \ref{fig:phd1}:
(a,b) $A$, $B$, $C$ and $D$ in phases I-IV, and
(c,d) $E$ and $F$ in phases V and VI.
Parameters: $\Delta=0.5$ and $L=8$ sites.}
\label{fig:Szz}
\end{figure}

The most important change at finite $\Delta$ occurs for the phase
transition between phases V and III (AF/AO) which becomes continuous for 
fixed $y$ and decreasing $x$, with a gradual change of spin correlations 
from the alternating singlets to an AF order along the chain \cite{You15}.
Here we discuss in more detail the intermediate case of $\Delta=0.5$.
First we address the phases with uniform spin-orbital order. The spin
structure factors has distinct peaks at $k=0$ for FM order and at
$k=\pi$ for AF order. Similarly, one finds a maximum of the orbital
structure factor $T^{zz}(k)$ at $k=0$ for FO order and at $k=\pi$ for
AO order. These structure factors complement one another and one finds
that the spin correlations are somewhat weaker due to stronger spin
fluctuations, while the orbital fluctuations are moderate at this value
of $\Delta=0.5$.

\begin{figure}[t!]
\begin{center}
\includegraphics[width=8.5cm]{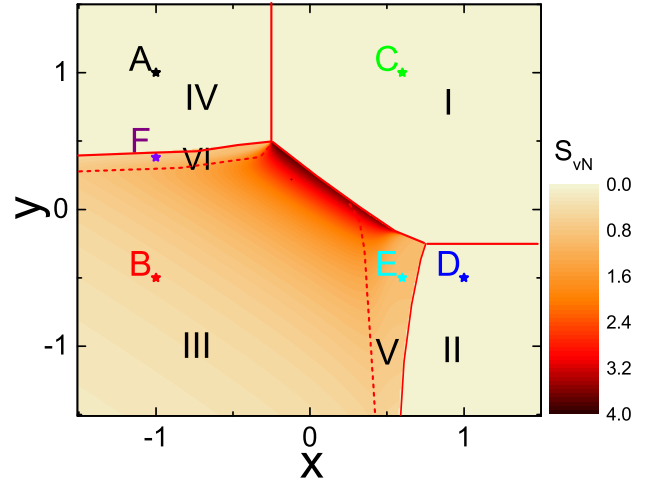}
\end{center}
\caption{(Color online)
Phase diagram in the $(x,y)$ plane and spin-orbital entanglement
${\cal S}_\textrm{vN}^0$ (\ref{svn}) (right scale) in different phases
(indicated by color intensity) of the anisotropic 
spin-orbital SU(2)$\otimes XXZ$ model (\ref{som}) with $\Delta=0.5$, 
obtained for a ring of $L=8$ sites. Phases I-IV correspond to FM/FO, 
AF/FO, AF/AO, FF/AO order in spin-orbital sectors.
Orbitals (spins) follow a quadrupled pattern accompanied by spin
(orbital) dimer correlations in phase V (VI). Six labeled points are:
$A=(-1,1)$, $B=(-1,-0.5)$, $C=(0.5,1)$, $D=(0.5,-0.5)$, $E=(1,-0.5)$,
and $F=(-1,0.38)$ --- they are used to investigate spin and orbital
structure factors in different phases, see Fig. \ref{fig:Szz}.
The phase boundaries, determined by the dominant modes of structure
factors are shown by solid (dashed) lines 
for the first (second) order quantum phase transitions.}
\label{fig:phd1}
\end{figure}

The dimerised phase V is characterized by a remarkably different
behavior, see Figs. \ref{fig:Szz}(c) and \ref{fig:Szz}(d). Phase V,
found at the $E$ point in Fig. \ref{fig:phd1}, has spin dimers 
accompanied by an orbital pattern with the periodicity of four sites, 
see Fig. \ref{fig:scheme}(a). Spin correlations give a sharp maximum of
$S^{zz}(k)$ at $k=\pi$ as for AF states, while two symmetric peaks of
$T^{zz}(k)$ at $k=\pi/2$ and $k=3\pi/2$ indicate quadrupling of the
unit cell in the orbital channel. When the model evolves towards the
SU(2)$\otimes$SU(2) limit with increasing $\Delta$, one expects also
a similar phase VI with interchanged role of spin and orbital
correlations. Indeed, this complementary phase emerges already at small
$\Delta>0$ and is identified by the respective structure factors shown 
also in Figs. \ref{fig:Szz}(c) and \ref{fig:Szz}(d).

\begin{table}[b!]
\caption{The spin-orbital configurations, momenta $k$, the total spin
$S^z$ and orbital $T^z$ quantum numbers of the ground states I-VI found
for the anisotropic SU(2)$\otimes XXZ$ model at $\Delta<1$.
All states in these subspaces are nondegenerate ($d=1$), but in case
of $S^z=L/2$ there are $L+1$ degenerate states for total $S=L/2$,
and for $T^z=L/2$ and $\Delta<1$ an equivalent state for $-$ orbitals
has $T^z=-L/2$. At $\Delta=0$ phase VI is absent and the ground state
degeneracy of phase V changes to $d=4$ corresponding to momenta
$k=0,\pm\pi/2,\pi$. }
\label{tab:phd}
\begin{ruledtabular}
\begin{tabular}{cccccc}
 phase & spin state & orbital state & $k$ & $S^z$ & $T^z$    \\
\hline
 I & $\uparrow\uparrow\uparrow\uparrow\uparrow\uparrow\uparrow\uparrow$
   &$+ + + + + + + +$&        0       & $L/2$ & $L/2$   \cr
II & $\uparrow\downarrow\uparrow\downarrow\uparrow\downarrow\uparrow\downarrow$
   &$+ + + + + + + +$&        0       &   0   & $L/2$   \cr
III& $\uparrow\downarrow\uparrow\downarrow\uparrow\downarrow\uparrow\downarrow$
   &$+ - + - + - + -$&        0       &   0   &   0     \cr
IV & $\uparrow\uparrow\uparrow\uparrow\uparrow\uparrow\uparrow\uparrow$
   &$+ - + - + - + -$&        0       & $L/2$ &   0     \cr
 V & ($S=0$) singlets
   &$+ - - + + - + -$&        0       &   0   &   0     \cr
VI & $\uparrow\downarrow\downarrow\uparrow\uparrow\downarrow\downarrow\uparrow$
   & $(T=0)$ singlets&        0       &   0   &   0     \cr
\end{tabular}
\end{ruledtabular}
\end{table}

The above analysis of the structure factors demonstrates that the phase
diagram found for $\Delta=0.5$ contains six distinct phases, see Fig.
\ref{fig:phd1}. The quantum phase transitions at the boarder lines I-V
and II-V are of first order. The phase transition between phases III 
(AF/AO) and V is a first order transition only for $\Delta=0$, and here 
this transition is continuous \cite{Gu07}. As described above, phase VI
emerges at finite $\Delta$ but is still quite narrow in the phase
diagram of Fig. \ref{fig:phd1}. Also the phase transition from phase
III to phase VI is continuous.
We have verified that due to the short-range nature of spin-orbital 
correlations, the size $L=12$ is sufficient as the phase boundaries 
are here almost the same as for the ring of $L=8$ sites. 

The spin-orbital phases I-VI found at $\Delta>0$ are summarized in
Table I. Phases I-IV
have polarized or alternating spin and orbital components, combined in
all possible ways into phases: FM/FO, AF/FO, AF/AO, and FM/AO. Phases
with either $S^z=L/2$ or $T^z=L/2$ (FM or FO) have of course also
degeneracy with respect to other possible values of $S^z$ or $T^z$
(the latter only at $\Delta=1$ when $T$ is also a good quantum number).
In addition, there are two phases with dimer orbital (phase V) or dimer
spin (phase VI) correlations. It is remarkable that these two phases
survive in the isotropic model at $\Delta=1$, see Sec. \ref{sec:xxx}.

\begin{figure}[b!]
\includegraphics[width=8.4cm]{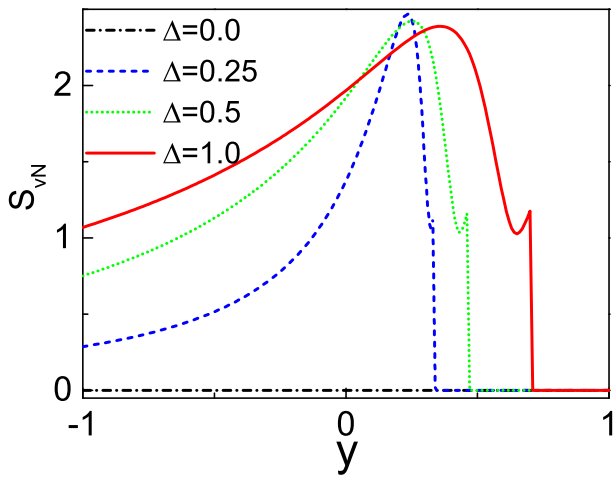}
\caption{(Color online) Spin-orbital entanglement entropy
${\cal S}^0_\textrm{vN}$ (\ref{som}) in the ground state of the
spin-orbital model (\ref{som}) as a function of $y$ for selected values
of $\Delta$. The onset of phase VI is detected at $\Delta>0$ by a
step-like increase of entropy to ${\cal S}^0_\textrm{vN}=1$.
Parameters: $x=-0.5$ and $L=8$.}
\label{fig:S_v}
\end{figure}

For $\Delta>0$ also phase III  is characterized by finite SOE, and it
expands to higher values of $y$ along vertical lines for fixed 
$x<-0.25$. Indeed, the onset of entangled region in the phase diagram
of Fig. \ref{fig:phd1} moves to higher values of $y$ with increasing 
$\Delta$, see Fig. \ref{fig:S_v}. At $\Delta=0.25$ the entanglement 
entropy ${\cal S}^0_\textrm{vN}$ develops a narrow peak with a maximum 
at $y\simeq 0.15$. This maximum broadens up and moves somewhat to the
right (to higher $y$) when the orbital fluctuations increase with 
increasing $\Delta$ towards the isotropic SU(2)$\otimes$SU(2) model. 
A sharp increase of the entropy to ${\cal S}^0_\textrm{vN}=1$, visible 
in all the curves shown in Fig. \ref{fig:S_v}, signals the SOE in phase 
VI which increases with increasing $\Delta$. 
This large SOE can develop because phase VI is similar to phase V --- 
it does not break translation invariance of the model and different 
spin-orbital configurations contribute simultaneously.

\begin{figure}[t!]
\includegraphics[width=8.4cm]{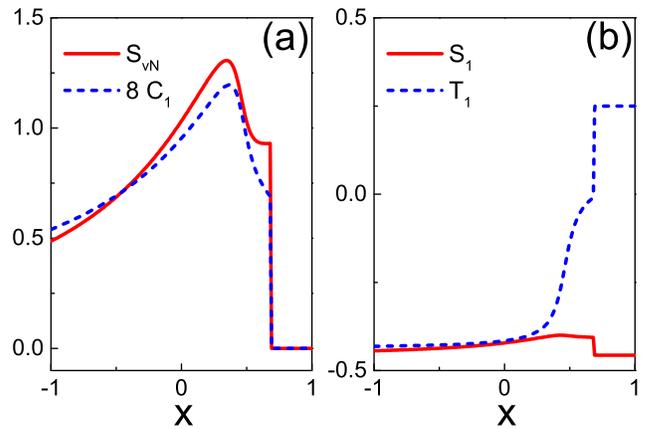}
\caption{(Color online) The onset of phase V and its gradual change
into phase III with decreasing $x$ in the ground state of the
spin-orbital model (\ref{som}):
(a) spin-orbital entanglement entropy ${\cal S}^0_\textrm{vN}$
(\ref{svn}) and the spin-orbital correlation function $C_1$ (\ref{cij});
(b) spin $S_1$ (\ref{sij}) and orbital $T_1$ (\ref{tij}) correlation
functions. The onset of phase V is signaled by a step-like increase of
the entropy to ${\cal S}^0_\textrm{vN}=1$.
Parameters: $\Delta=0.5$, $y=-0.5$ and $L=8$.}
\label{fig:S_y}
\end{figure}

To detect SOE we employ here not only the vNE,
${\cal S}^0_\textrm{vN}$ Eq. (\ref{svn}), but also a direct measure by
the spin-orbital correlation function on a bond \cite{Ole06},
\begin{equation}
\label{cij}
C_{1}\!\equiv\frac{1}{L}\sum_{i=1}^L
\Big[\Big\langle\!({\vec S}_i\cdot{\vec S}_{i+1})
                ({\vec T}_i\cdot{\vec T}_{i+1})\!\Big\rangle
     -\Big\langle{\vec S}_i\cdot{\vec S}_{i+1}\Big\rangle
      \Big\langle{\vec T}_i\cdot{\vec T}_{i+1}\Big\rangle\Big],
\end{equation}
and we compare it with the conventional intersite spin- and orbital
correlation functions:
\begin{eqnarray}
\label{sij}
S_{1}&\equiv&\frac{1}{L}\sum_{i=1}^L
\Big\langle{\vec S}_i\cdot {\vec S}_{i+1}\Big\rangle, \\
\label{tij}
T_{1}&\equiv& \frac{1}{L}\sum_{i=1}^L
\Big\langle{\vec T}_i\cdot {\vec T}_{i+1}\Big\rangle.
\end{eqnarray}
The above general expressions imply averaging over the exact
(translation invariant) ground state found from Lanczos
diagonalization of a ring of length $L$. While $S_{1}$ (\ref{sij}) 
and $T_{1}$ (\ref{tij}) correlations indicate the tendency towards 
particular spin and orbital order, $C_{1}$ (\ref{cij}) quantifies the 
SOE --- if $C_{1}\ne 0$ spin and
orbital degrees of freedom are entangled and the mean-field decoupling
cannot be applied in Eq. (\ref{som}) as it generates systematic errors.

To gain a better insight into the nature of a phase transition between
phases II (AF/FO) and V and between V and III (AF/AO) which occur for 
decreasing $x$ at a fixed $y<-1/4$, we study SOE vNE 
${\cal S}^0_\textrm{vN}$, joint spin-orbital $C_1$ (\ref{cij}), and 
individual spin $S_1$ (\ref{sij}) and orbital $T_1$ (\ref{tij}) 
correlations in Fig. \ref{fig:S_y}. Both ${\cal S}^0_\textrm{vN}$ and 
$C_1$ show a very similar behavior with a maximum within phase V, see 
Fig. \ref{fig:S_y}(a). The SOE is lower in phase III than in phase V, 
i.e., below $x\simeq-0.25$, and decreases further with decreasing $x$. 
These two phases have rather similar spin correlations $S_1$, but 
orbital correlations $T_1$ are similar to spin ones only within phase 
III (AF/AO); above $x\simeq -0.25$ they
vary fast within phase V and almost disappear ($T_1\simeq 0$) near the
transition point to phase II (AF/FO), see Fig. \ref{fig:S_y}(b).

\subsection{Isotropic spin-orbital SU(2)$\otimes$SU(2) model}
\label{sec:xxx}

\begin{figure}[t!]
\hskip -.4cm
\begin{center}
\includegraphics[width=8.4cm]{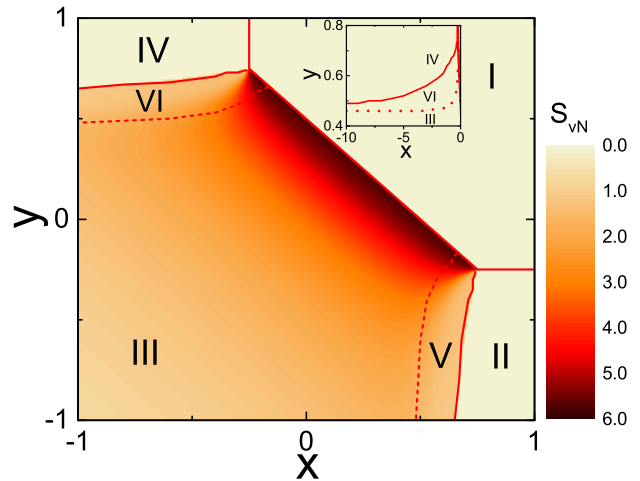}
\end{center}
\caption{(Color online)
Phase diagram in the $(x,y)$ plane and spin-orbital entanglement
${\cal S}_\textrm{vN}^0$ (\ref{svn}) (right scale) in the
ground state of the isotropic spin-orbital SU(2)$\otimes$SU(2) model
(\ref{som}) with $\Delta=1.0$ obtained for a ring of $L=12$ sites.
Phases I-IV correspond to FM/FO, AF/FO, AF/AO, FM/AO order in
spin-orbital sectors. Orbitals (spins) follow a quadrupled pattern
accompanied by spin (orbital) dimer correlations in phase V (VI). The
phase boundaries determined by dominant modes of structure factors,
shown by solid (dashed) lines, are of first (second) order.
Inset shows the extended range of phase VI which separates phases IV
and III for $-10<x<0$.}
\label{fig:phd2}
\end{figure}

The phase diagram of the isotropic SU(2)$\otimes$SU(2) model (Fig.
\ref{fig:phd2}) includes the same six phases as the one obtained at
$\Delta=0.5$ (Fig. \ref{fig:phd1}), with spin-orbital correlations
explained in Table I. The main differences to the phase diagram at
$\Delta=0.5$ is a somewhat reduced stability range of phase IV (FM/AO), 
and also phase II (AF/FO), both destabilized by enhanced spin-orbital 
fluctuations in and around phase III (AF/AO). The range of entangled 
ground states is broad and includes phase III as well as phases V and 
VI on its both sides. The phase transitions from phase I where all 
quantum fluctuations are absent to either phase II or IV are given by 
straight lines and may be determined using mean-field approach.

It is quite unexpected that the dimerised phases V and VI survive in
the phase diagram of the isotropic SU(2)$\otimes$SU(2) model at
$\Delta=1.0$, see Fig. \ref{fig:phd2}. These two phases emerge in
between a disentangled phase II and III in the case of phase V, and 
similarly between phases IV and III in the case of phase VI, and are 
stabilized by robust quadrupling of orbital or spin correlations which 
was overlooked before \cite{You12}. 
This is not so surprising as one expects that isotropic spin-orbital 
interactions would lead instead to uniform phases only. In each case 
the effective exchange interaction changes sign in one (either spin 
or orbital) channel which resembles the mechanism of exotic magnetic 
order found in the Kugel-Khomskii model \cite{Brz12}.

\begin{figure}[b!]
\begin{center}
\includegraphics[width=\columnwidth]{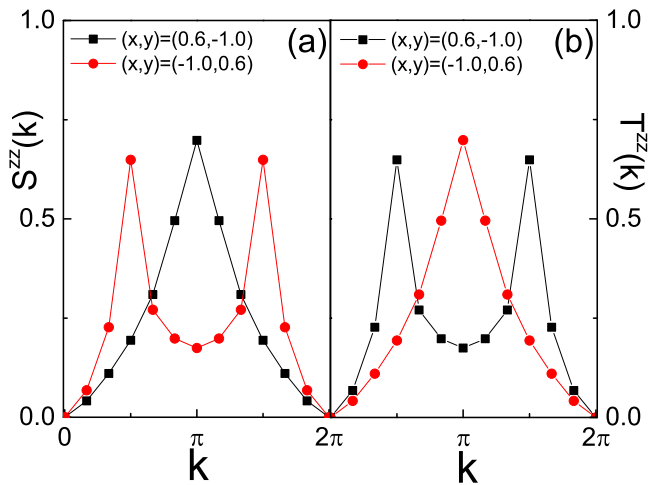}
\end{center}
\caption{(Color online) The structure factors obtained for a 
spin-orbital ring Eq. (\ref{som}) of $L=8$ sites in the full Hilbert 
space at $\Delta=1.0$:
(a) spin $S^{zz}(k)$, and
(b) orbital $T^{zz}(k)$. 
The points $(0.6,-1.0)$ (filled squares) and $(-1.0,0.6)$ 
(filled circles) correspond to phase V and VI, see Fig. \ref{fig:phd2}.
}
\label{fig:SU(2)}
\end{figure}

The phases V and VI emerge by the same mechanism as phase V for the
Ising orbital interactions, see Sec. \ref{sec:ising}. In the isotropic
model this phase and phase VI with complementary spin-orbital
correlations occur in a symmetric way with respect to the $x=y$ line,
see Fig. \ref{fig:phd2}. Orbital correlations in the case of phase V
(spin correlations in the case of phase VI) change gradually towards FO
(FM) order in phase II (IV) with increasing $x$ ($y$). The quadrupling 
of the unit cell seen in both phases in such correlations, shown in Fig. 
\ref{fig:SU(2)}, may be seen as a precursor of this transition. Both 
phases are stable only in a rather narrow range and disappear for 
$y\to-\infty$ or $x\to-\infty$, respectively, as presented in the inset 
of Fig. \ref{fig:phd2} for phase VI.
The spin-orbital interactions and the mechanism stabilizing these phases
are different from spin-Peierls and orbital-Peierls mechanisms in the
1D SU(2)$\otimes$SU(2) spin-orbital model with positive exchange
($J=-1$) \cite{Li05}. We emphasize that the present mechanism of
dimerization is effective only in one (spin or orbital) channel and thus
it is also distinct from the spin-orbital dimerization in a FM chain 
found at finite temperature \cite{Sir08}.

\begin{figure}[t!]
\includegraphics[width=8.4cm]{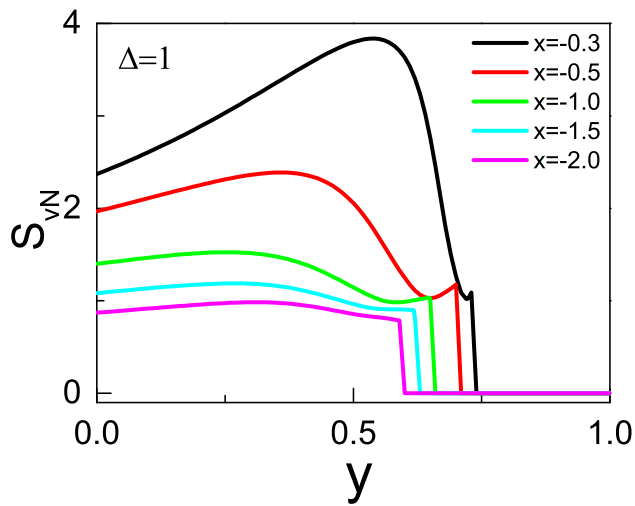}
\caption{(Color online) Spin-orbital entanglement entropy
${\cal S}^0_\textrm{vN}$ in the ground state of the SU(2)$\otimes$SU(2)
spin-orbital model (\ref{som}) as a function of $y$ for representative
values of $x\in[-2.0,-0.3]$.
The onset of phase VI at decreasing $y$ is detected by a step-like
increase of entropy at the IV-VI phase transition.}
\label{fig:allx}
\end{figure}

A characteristic feature of SOE in phase VI (and similar in phase V)
is a competition between the spin (orbital) quadrupling correlations
along the chain which support orbital (spin) singlets, with the AF/AO
order characteristic of phase III (AF/AO). For $x\in[-1.0,-0.3]$ the 
vNE ${\cal S}^0_\textrm{vN}$ increases discontinuously at the phase 
transition IV-VI, next drops somewhat and next increases further, see 
Fig. \ref{fig:allx}. It exhibits a broad maximum, moving to lower 
values of $y$ with decreasing $x$. This behavior shows that two phases 
(VI and III) compete in this regime. For still lower values of $x$ the 
SOE entropy is smaller and almost constant when $y$ decreases deeply 
into phase III.

\subsection{Entanglement in the SU(2)$\otimes$SU(2) models}
\label{sec:III}

The ground state obtained in the present spin-orbital 
SU(2)$\otimes$SU(2) model for phase III (AF/AO) is distinct from the one 
found for positive coupling constant, i.e., $J=-1$ in Eq. (\ref{som}). 
We elucidate this difference by studying both models along the symmetry
line $x=y$ in the phase diagram. Phase I (FM/FO) is found in the present 
case for $x=y>1/4$, while in the case of positive coupling constant it 
becomes the ground state for $x=y<-1/4$ \cite{Ori00}.

\begin{figure}[t!]
\begin{center}
\includegraphics[width=8.4cm]{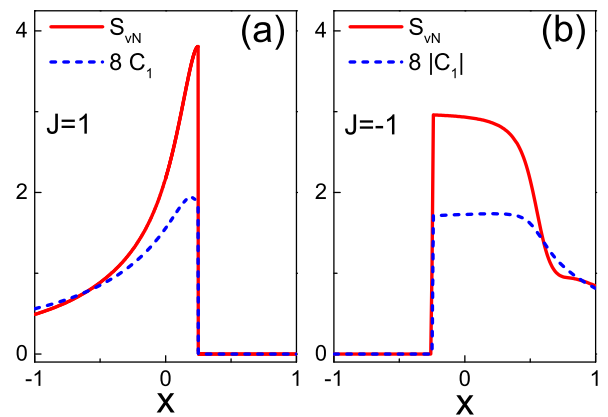}
\end{center}
\caption{(Color online) von Neumann entropy ${\cal S}_\textrm{vN}^0$
(\ref{svn}) and joint spin-orbital bond correlation $C_1$ (\ref{cij})
as obtained in the SU(2)$\otimes$SU(2) model (\ref{som}) along the
symmetry $x=y$ line in the phase diagram with the ring of $L=8$ sites
for:
(a) the model with negative coupling $-J=-1$ \cite{You12}, and
(b) the model with positive coupling $-J=1$ constant \cite{Li99}.
}
\label{fig:y=x}
\end{figure}

First we consider SOE detected by the vNE
${\cal S}^0_\textrm{vN}$ Eq. (\ref{svn}), and by the joint spin-orbital
correlation function $C_1$ (\ref{cij}), see Fig. \ref{fig:y=x}. To
understand better the transition from phase I to phase III, we consider
the correlation functions along the symmetry line $x=y$. Phase I with
FM/FO order is disentangled in both cases. At the quantum phase
transition to phase III, signalled by a rapid increase of both
${\cal S}_\textrm{vN}^0$ and $8|C_1|$, we observe that the entropy
reaches the highest value at the onset of phase III, and then decreases
when $x$ decreases and one moves deeper into the entangled phase III.
In the present model (\ref{som}) one finds $C_1>0$ which is imposed by
the negative coupling constant. The entropy maximum and also the
maximum of $8C_1$ are sharp indeed and signal the onset of phase III,
see Fig. \ref{fig:y=x}(a). The model with a positive coupling constant
behaves differently --- here the joint spin-orbital correlations are
negative $C_1<0$, and both ${\cal S}_\textrm{vN}^0$ and $8|C_1|$ have
flat maxima in a range of $x$ and only at $x\simeq 0.5$ both drop 
rapidly. This large SOE for $x\in[-0.25,0.5]$ indicates a spin-orbital 
liquid phase which forms near the SU(4) point $x=y=1/4$ \cite{Ori00}. 
Only at $x>0.5$ the strong spin-orbital fluctuations are weakened when 
phase III is approached and SOE decreases.

A special feature of the present SU(2)$\otimes$SU(2) spin-orbital model
is a very distinct behavior along the I-III phase transition line
$x+y=1/2$, see Fig. \ref{fig:phd2}. The ground state energy of phase I
(FM/FO) in which quantum fluctuations are absent, $E^0=-J(1/4+x)^2$, is 
found by taking exact classical values of spin (\ref{sij}) and orbital 
(\ref{tij}) correlations on the bonds, $S_1=T_1=1/4$. The energy 
decreases when the transition at $x=1/4$ is approached. On the 
contrary, coming from the other side it is not allowed to assume
classical correlations, $S_1=T_1=-1/4$, as then the Hamiltonian
would vanish at the transition. In fact the energy $E_0=-J/4$ can
be also obtained mainly from enhanced joint spin-orbital correlations,
$\langle(\vec{S}_{j}\cdot\vec{S}_{j+1})
(\vec{T}_{j}\cdot\vec{T}_{j+1})\rangle=5/16$ and $C_1=1/4$,
see Fig. \ref{fig:y=x}. At the phase transition the spin and orbital
correlations are \textit{very weak}, i.e., $S_1=T_1\simeq -1/8$.
This is a very peculiar situation as joint spin-orbital correlations
cannot be factorized and damp to a large extent individual spin and
orbital correlations.

\begin{figure}[t!]
\begin{center}
\includegraphics[width=8.4cm]{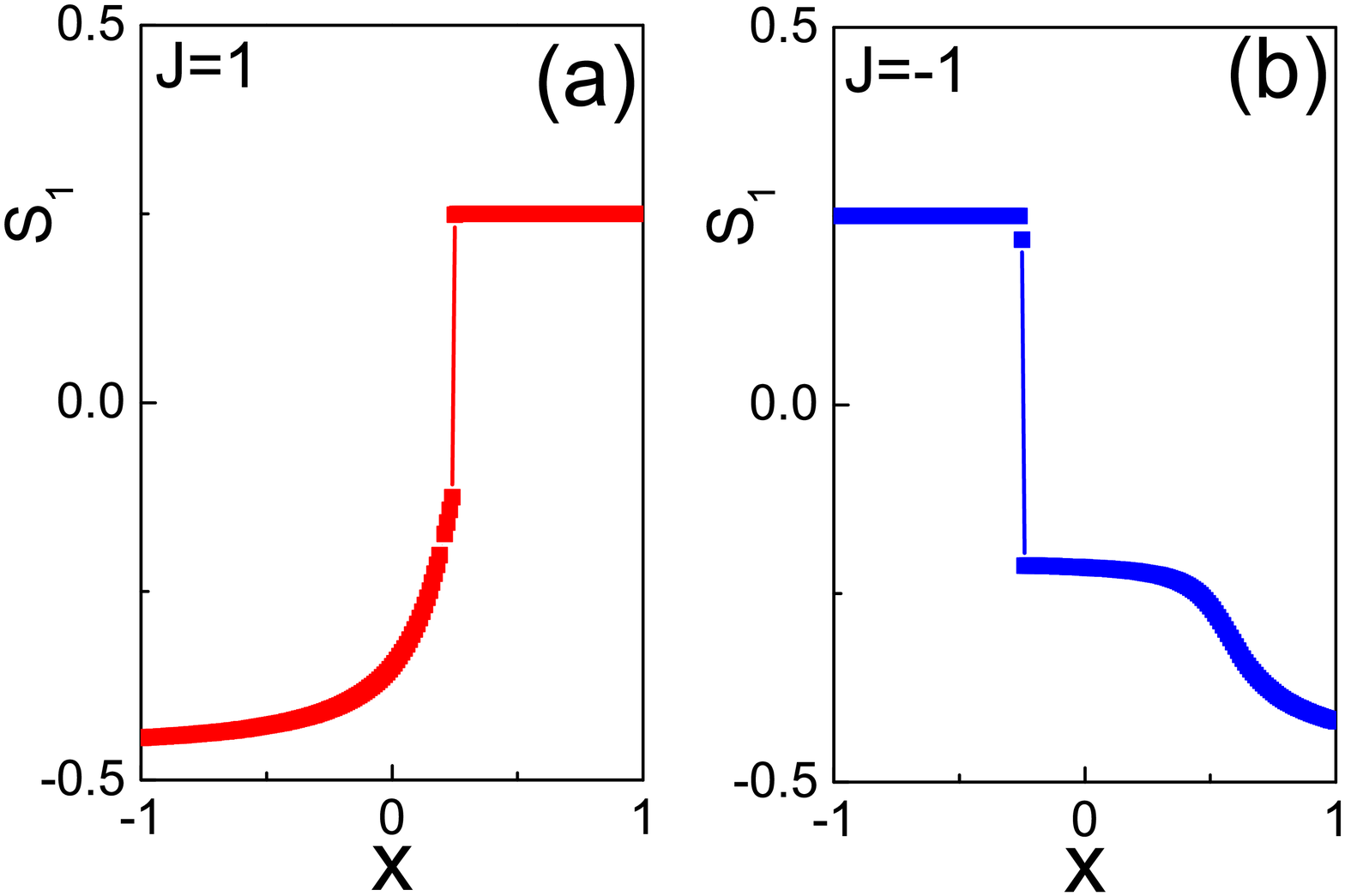}
\end{center}
\caption{(Color online) Spin $S_1$ (\ref{sij}) and orbital $T_1$ 
(\ref{tij}) ($T_1=S_1$) bond correlations as obtained in the 
SU(2)$\otimes$SU(2) model (\ref{som}) along the symmetry line $x=y$ in 
the phase diagram with the ring of $L=8$ sites for:
(a) the model with negative exchange $J=1$ \cite{You12}, and
(b) the model with positive exchange $J=-1$ \cite{Li99}.
}
\label{fig:sandt}
\end{figure}

\begin{figure}[b!]
\begin{center}
\includegraphics[width=8.4cm]{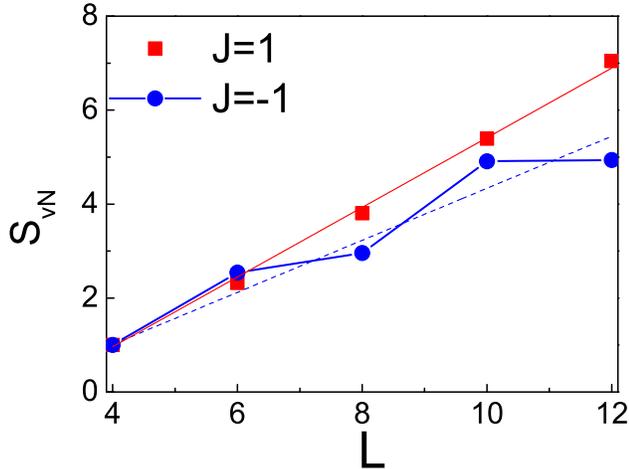}
\end{center}
\caption{(Color online) von Neumann entropy ${\cal S}_\textrm{vN}^0$
(\ref{svn}) obtained in the isotropic SU(2)$\otimes$SU(2) model 
(\ref{som}) at the maximum seen in Fig. \ref{fig:y=x}(a) at $x=y=0.249$ 
(squares) and at $J=-1$ at $x=y=-0.249$ (circles), i.e., at the onset 
of phase III, both for rings of $L=4,6,\cdots,12$ sites and the linear 
fits to the data (dashed lines).
}
\label{fig:lin_S}
\end{figure}

A qualitatively different nature of the SOE in both SU(2)$\otimes$SU(2)
models is also captured by its effect on the individual spin (orbital)
correlations, see Fig. \ref{fig:sandt}. In the present model 
(\ref{som}) with $J=1$, $C_1>0$ damps individual spin and orbital 
fluctuations near the quantum phase transition stronger than in the 
case of positive coupling ($J=-1$), cf. Figs. \ref{fig:sandt}(a) and 
\ref{fig:sandt}(b). However, the joint fluctuations $C_1$ decrease fast
when $J=1$, while they are robust in the spin liquid phase for $J=-1$
when $x\in[0.25,0.50)$. At $x=0$ one finds $S_1=T_1\simeq-0.37$, 
already much below the value of $S_1=T_1\simeq-0.22$ found for $J=-1$. 
The spin $S_1$ and orbital $T_1$ correlations become similar in both 
phases deeply within phase III, as found by comparing these values at 
$x=-1$ for $J=1$ and at $x=1$ for $J=-1$. Here SOE is weak because 
spins and orbitals fluctuate almost independently.

The sharp peak of ${\cal S}_\textrm{vN}^0$ found near the phase
transition for $J=1$ is rather unusual, see Fig. \ref{fig:y=x}(a). In
this regime of parameters the ground state energy $E_0$ is lowered by
positive joint spin-orbital correlations $C_1$, while negative
$S_1=T_1\simeq -1/8$ [Fig. \ref{fig:sandt}(a)] increase it somewhat.
Large vNE ${\cal S}_\textrm{vN}^0$ is found near the phase transition 
for rings of even length, starting from
${\cal S}_\textrm{vN}^0\simeq 1$ for $L=4$. It is remarkable that the 
vNE at the maximum scales with system size $L$, see Fig. 
\ref{fig:lin_S}. This behavior is unique and proves that SOE 
which takes place at every bond is extensive and 
extends here over the entire ring. The model with positive coupling
constant, $J=-1$, has also a similar linear scaling,
${\cal S}_\textrm{vN}^0\propto L$, but the entropy is smaller and
systematic fluctuations between the rings of length of $4n$ and $4n+2$ 
sites, seen in Fig. \ref{fig:lin_S}. are distinct and indicate a
crucial role played here by global SU(4) singlets.

\section{Entangled elementary excitations: FM/FO order}
\label{sec:ff}

\subsection{Analytic approach}
\label{sec:ana}

When the ground state is disentangled, SOE is generated locally in
excited states \cite{Plaq} and would not scale linearly with system
size. Indeed, we analyzed the low-energy excitations of the
disentangled FM/FO phase in the 1D SU(2)$\otimes$SU(2) spin-orbital
model in Ref. \cite{You12} and found a much weaker dependence on system
size. We investigated the SOE for spin-orbital bound states (BSs) and
spin-orbital exciton (SOEX) state and found a logarithmic scaling,
while the entropy saturates for other separable (trivial) spin-orbital
excitations. One finds that the vNE is controlled by the spin-orbital
correlation length $\xi$ and decays logarithmically with ring length 
$L$.

Here we consider again the FM/FO disentangled ground state
$\vert 0\rangle$, obtained for the anisotropic spin-orbital 
SU(2)$\otimes XXZ$ model with the exactly known ground state energy 
$E^0$,
\begin{eqnarray}
H \vert 0\rangle = E^0 \vert 0\rangle.
\end{eqnarray}
Using equation of motion method one finds spin (magnon) excitations
with dispersion
\begin{equation}
\omega_S(Q)=  \left(\frac{1}{4}+y\right)(1-\cos Q),
\label{spinexcitations}
\end{equation}
and orbital (orbiton) excitations \cite{Herzog},
\begin{equation}
\omega_T(Q)=  \left(\frac{1}{4}+x\right)(1-\Delta \cos Q).\label{orbex}
\end{equation}
The spin-orbital continuum is given by
\begin{equation}
\label{mode}
\Omega(Q,q)=\omega_S\left(\frac{Q}{2}-q\right)
           +\omega_T\left(\frac{Q}{2}+q\right).
\end{equation}

Next, we consider the propagation of a magnon-orbiton pair excitation
along the FM/FO chain, by exciting simultaneously a single spin and a 
single orbital. The translation symmetry imposes that total momentum 
$Q=2m\pi/L$ (for $m=0,\cdots,L-1$) is conserved during scattering.
The scattering of magnon and orbiton with initial (final) momenta
$\{\frac{Q}{2}-q,\frac{Q}{2}+q\}$ ($\{\frac{Q}{2}-q',\frac{Q}{2}+q'\}$)
and the total momentum $Q$ is represented by the Green's function
\cite{Wor63},
\begin{equation}
G(Q,\omega)= \frac{1}{L}\sum_{q,q'}\;\langle\langle
S_{\frac{Q}{2}-q'}^+T_{\frac{Q}{2}+q'}^+ | 
S_{\frac{Q}{2}-q}^-T_{\frac{Q}{2}+q}^-\rangle\rangle,
\label{Greenfunction}
\end{equation}
for a combined spin ($S_{\frac{Q}{2}-q}^-$) and orbital
($T_{\frac{Q}{2}+q}^-$) excitation.  The analytical form reads
\begin{eqnarray}
\!\!G(Q,\omega) &=&
G^0(Q,\omega) + \Pi(Q,\omega),  \label{GqqDeltale1} \\
\!\!\Pi(Q,\omega)&=&
- \frac{2(1+\Delta)+(1-\Delta^2) F_{ss}(Q,\omega)}{4[1
+\Lambda(Q,\omega)]}\,H_{cc}^2(Q,\omega)
\nonumber \\
&-& \frac{2  (1-\Delta) +  (1-\Delta^2) F_{cc}(Q,\omega)}
{4[1+\Lambda(Q,\omega)]}\,H_{ss}^2 (Q,\omega)     \nonumber \\
&+&\frac{(1-\Delta^2)F_{sc}(Q,\omega)}{2[1+\Lambda(Q,\omega)]}
H_{cc}(Q,\omega)\, H_{ss}(Q,\omega),  \label{piq}
\end{eqnarray}
where the noninteracting Green's function is given by
\begin{eqnarray}
G^0(Q,\omega) &=& \frac{1}{L} \sum_q G_{qq}^0 (Q,\omega), \\
G_{qq}^0 (Q,\omega) &=& \frac{1}{\omega-\Omega(Q,q)}\,,
\end{eqnarray}
and
\begin{eqnarray}
\!\!\!\!\!\!\!\!H_{cc} (Q,\omega)&=& \frac{1}{L} \sum_{q}
\left(\cos \frac{Q}{2} -\cos q \right) G_{qq}^0(Q,\omega), \\
\!\!\!\!\!\!\!\!H_{ss} (Q,\omega)&=& \frac{1}{L} \sum_{q}
\left(\sin \frac{Q}{2} -\sin q \right) G_{qq}^0(Q,\omega).
\end{eqnarray}
One finds the denominator in $\Pi(Q,\omega)$ (\ref{piq}),
\begin{eqnarray}
&&1+\Lambda(Q,\omega) \nonumber \\
&=&\left[1+\frac{1}{2}(1+\Delta)
F_{cc}(Q,\omega)\right]\left[1+\frac{1}{2}(1-\Delta)
F_{ss}(Q,\omega)\right]\nonumber \\
&-&\frac{1}{4} (1-\Delta^2)
F_{sc}^2(Q,\omega)\,, \label{DenominatorofGreen}
\end{eqnarray}
with
\begin{eqnarray}
\label{Fcc}
F_{cc}(Q,\omega ) &=& \frac{1}{L} \sum_{q}
\frac{(\cos\frac{Q}{2}  - \cos q)^2}{\omega-\Omega(Q,q)},  \\
\label{Fss}
F_{ss}(Q,\omega ) &=& \frac{1}{L} \sum_{q}
\frac{(\sin\frac{Q}{2}  - \sin q)^2}{\omega-\Omega(Q,q)},  \\
\label{Fsc}
F_{sc}(Q,\omega ) &=& \frac{1}{L} \sum_{q}
\frac{(\sin\frac{Q}{2}  - \sin q) (\cos\frac{Q}{2}-\cos q)}
{\omega - \Omega(Q,q)}. \nonumber \\
\end{eqnarray}
Here we define a phase $\delta\in[0,2\pi]$ which quantifies the
difference of dynamic properties of magnon and orbiton excitations
throughout the Brillouin zone in the form
\begin{eqnarray}
\tan \delta=\frac{  (4y+1) -(4x+1)\Delta   }
{ (4y+1)+ (4x+1)\Delta }\,\tan\left(\frac{Q}{2}\right).
\label{delta}
\end{eqnarray}
For the symmetry line $x=y$ Eq. (\ref{delta}) greatly simplifies for 
the isotropic model at $\Delta=1$ and gives $\delta(Q)=0$, which has 
been studied in Ref. \cite{You12}. Also, a quantity describes the 
position relative to the continuum is given by
\begin{equation}
a(Q,\omega)= \frac{\omega-(x+y+\frac{1}{2}) }{b(Q)}\,,
\label{defination3}
\end{equation}
with
\begin{eqnarray}
b(Q)&=&\left[\left(x+\frac{1}{4}\right)^2\Delta^2
+\left(y+\frac{1}{4}\right)^2\right. \nonumber \\
&+&\left. 2\Delta\left(x+\frac{1}{4}\right)
\left(y+\frac{1}{4}\right)\cos Q \right]^{1/2}.
\end{eqnarray}
The excitations at the top and at the bottom of the continuum correspond
to $a(Q,\omega)= 1$ and $-1$, respectively. In the noninteracting case,
we have the imaginary and real parts of Eq. (\ref{GqqDeltale1}),
\begin{eqnarray}
\label{G0singular}
\Im G^{0}(Q,\omega)&=&
-\frac{\theta(1-\vert a(Q,\omega) \vert)}{b(Q)\sqrt{1-a^2(Q,\omega)}}\,, \\
\Re G^{0}(Q,\omega) &=&
\frac{\theta(a(Q,\omega)-1)}{b(Q)\sqrt{a^2(Q,\omega)-1}}   \nonumber  \\
&-&\frac{\theta(-1-a(Q,\omega))}{b(Q)\sqrt{a^2(Q,\omega)-1}}\,, 
\end{eqnarray}
where $\theta (x)$ is Heaviside step function whose value is zero for
negative argument and 1 for nonnegative argument. The frequency
dependence of the imaginary part exhibits square-root singularities
in Eq. (\ref{G0singular}) at the bottom and at the top of the 
continuum \cite{Sch81}.

\subsection{Numerical studies}
\label{sec:num}

We note that the inclusion of the spin-orbital attraction will smear out
the singularities by Eq. (\ref{GqqDeltale1}) since a more pronounced
divergence of the numerator than the denominator occurs when 
$a(Q,\omega)\to\pm 1$, and they cancel each other, i.e.,
$\Im G^{}(Q,\omega)=0$. Furthermore, the poles of $G(Q,\omega)$ are
determined by
\begin{equation}
1+\Lambda(Q,\omega)=0.
\label{bs}
\end{equation}

\begin{figure}[t!]
\includegraphics[width=8.2cm]{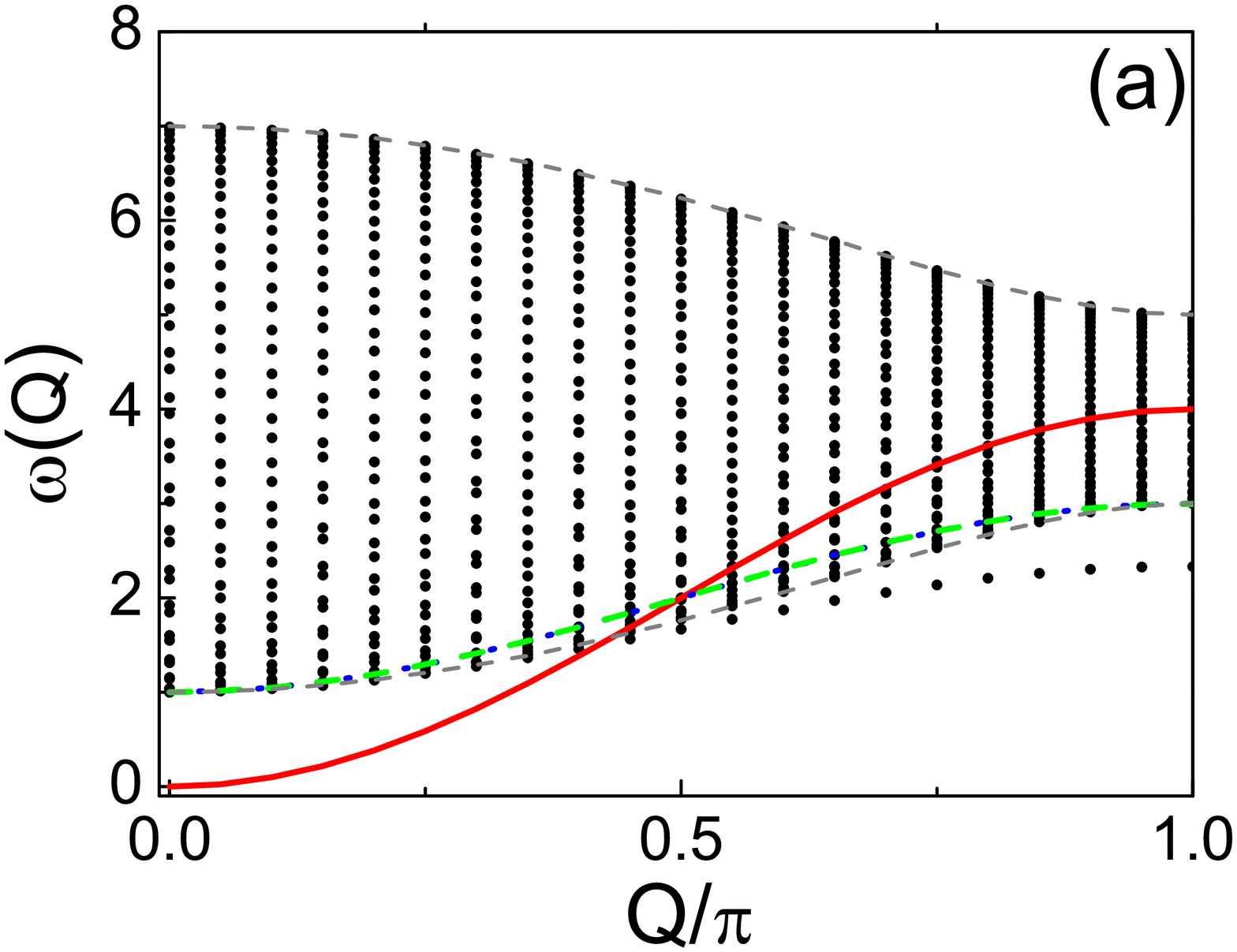}
\includegraphics[width=8.2cm]{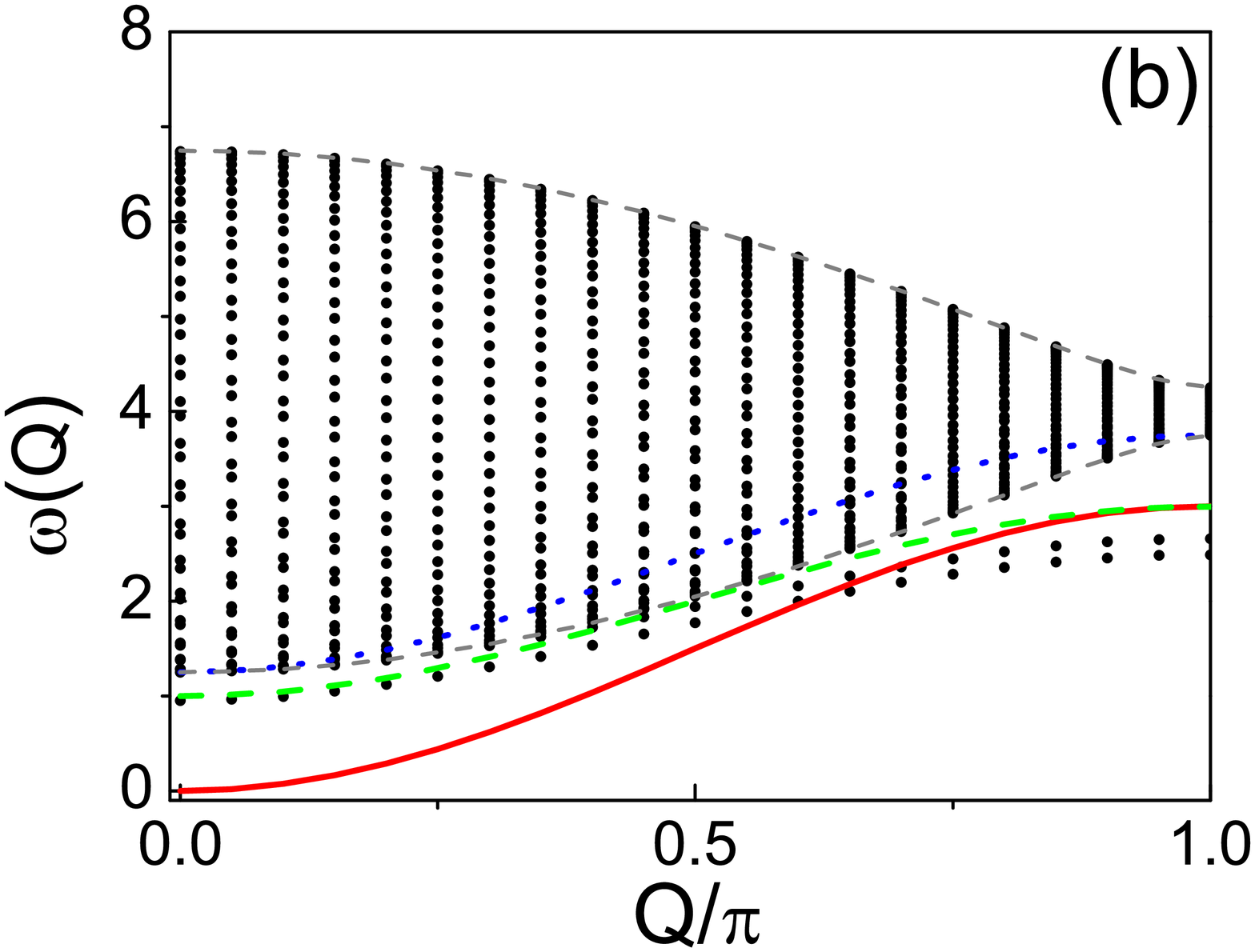}
\includegraphics[width=8.2cm]{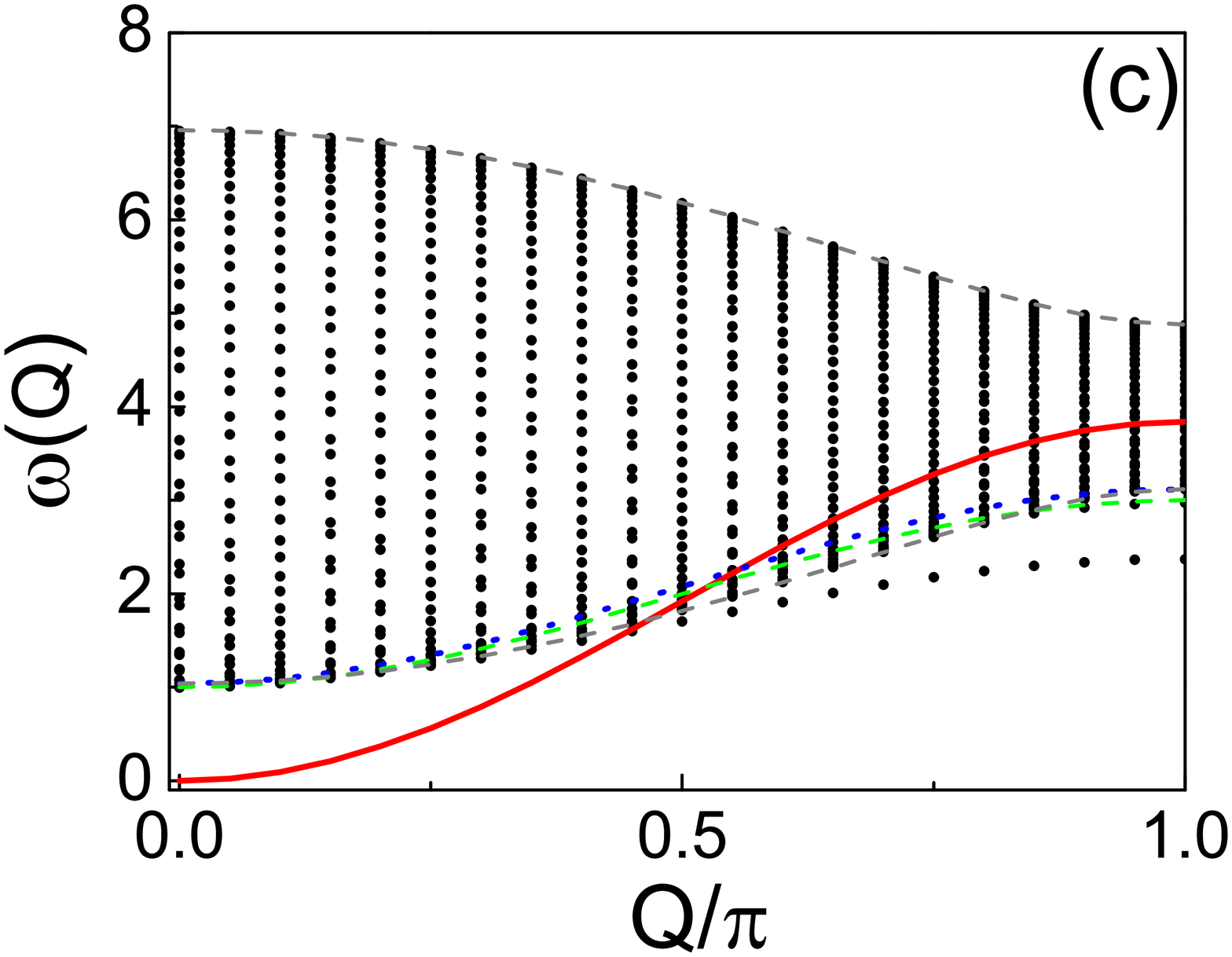}
\caption{(Color online)(a) Excitation spectra of a ring of $L=40$ sites
for the spin-orbital SU(2)$\otimes XXZ$ model with $\Delta=0.5$ as
function of momentum $Q$ at:
(a) $x=y=1/4$,
(b)~$x=0.375$, $y=0.125$, and
(c) $x=0.27$, $y=0.23$.
The dotted (blue), dashed (green) lines inside the spin-orbital
continuum $\Omega(Q,q)$ denote the orbital and SOEX excitations, i.e.,
$\omega_T(Q)$ and $\omega_\textrm{SOEX}(Q)$, respectively, that are
degenerate. The (red) solid lines show spin excitations. }
\label{fig:spectra}
\end{figure}

Our analysis shows that for given $Q$ most the real solutions of Eq.
(\ref{bs}) are interspersed within the continuum, but these modes are
unstable to two free waves. However, a small number of solutions may
lie well below the continuum. To investigate the spectra we begin with
the asymmetric SU(2)$\otimes XXZ$ model. As it is shown in Fig.
\ref{fig:spectra}, the attractive interactions shift spin-orbital BSs
outside the continuum \cite{Brink,Bal01,Coj03}. The binding energy
approaches zero for the isotropic SU(2)$\otimes$SU(2) model, but is
finite for anisotropic SU(2)$\otimes XXZ$ model due to a gap in the
orbital excitation spectrum, and the small-$q$ behavior of the binding
energy reveals that the BSs appear for arbitrarily small wave number.

\begin{figure}[t!]
\includegraphics[width=8.2cm]{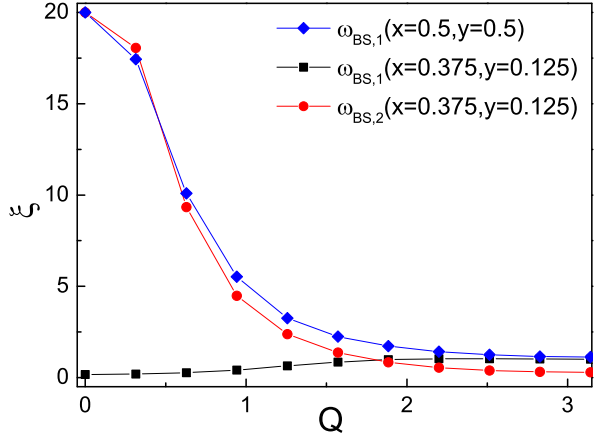}
\caption{(Color online) The correlation length $\xi$ for increasing
momentum $Q$ as obtained for a ring of $L=80$ sites in the anisotropic
SU(2)$\otimes XXZ$ model with $\Delta=0.5$.  }
\label{fig:xi}
\end{figure}

The BSs can be also obtained by the equation of motion method for a 
spin-orbital joint excitation, $S_m^-T_{m+l}^-\vert 0\rangle$. 
The collective mode follows from Eq. (\ref{bs}). Such a collective 
spin-orbital excitation (bound state) involves spin and orbital flips 
at many sites and can be written as follows,
\begin{eqnarray}
|\Psi(Q)\rangle &=& \frac{1}{\sqrt{L}} \sum_{m,l} a_l(Q) e^{iQm} S_{m}^-
T_{m+l}^-  \vert 0 \rangle \nonumber \\
&=&\sum_{q} a_q S_{\frac{Q}{2}-q}^-
T_{\frac{Q}{2}+q}^- \vert 0 \rangle,  \label{wavefunction}
\end{eqnarray}
with the coefficients
\begin{equation}
a_q =\frac{1}{\sqrt{L}}\sum_l a_l(Q)\,e^{-i\left(\frac{Q}{2}-q\right)l}.
 \label{alq}
\end{equation}
The correlation length $\xi\equiv\sum_l l\vert a_l\vert^2$ defines the 
average size of spin-orbital BSs or excitons and is much smaller than 
the system size, i.e., $0<\xi\ll L$. The correlation length becomes 
extensive for a trivial continuum state, as shown in Fig. \ref{fig:xi}.
The analytic solution of this equation is tedious but straightforward.
The dispersion of the collective excitation, $\omega_{\textrm{BS}}(Q)$,
can be analyzed in a simple way at some special points, including $Q=0$
and $Q=\pi$.

In the isotropic model (at $\Delta=1$), Eq. (\ref{bs}) reduces to
$1+F_{cc}(Q,\omega)=0$. In this case there is at most one solution for
every $Q$ \cite{You12}. Nevertheless, the anisotropic orbital coupling
will induce more branches in part of Brillouin zone (see Fig.
\ref{fig:spectra}). When $\Delta<1$ and $Q=0$, $F_{sc}(0,\omega)$=0 and
then Eq. (\ref{bs}) reduces to
\begin{eqnarray}
1+\frac{1}{2}(1+\Delta) F_{cc}(0,\omega)=0,
\label{BSfromFcc}
\end{eqnarray}
or
\begin{eqnarray}
1+\frac{1}{2}(1-\Delta) F_{ss}(0,\omega)=0.
\label{BSfromFss}
\end{eqnarray}
The solution that follows from Eq. (\ref{BSfromFcc}) is given by
\begin{eqnarray}
\omega_{\textrm{BS},1}(0) &=&  \frac{(\alpha +3 \beta)^2
-\sqrt{(\alpha-17\beta)(\alpha-\beta)^3}}{8 \beta} \nonumber \\
&+&x(1-\Delta)-3\beta+\frac{1}{2},
\label{1stBSk=0}
\end{eqnarray}
where $\alpha=(\Delta x +y)$, $\beta=(1+\Delta )/4 $. The instability of 
such a mode given by $\omega_{\textrm{BS},1}(0)=0$ sets up the threshold 
of the FM/FO state separating it from the AF/AO state (phase III)
\cite{Bal01,You12}. Moreover, the solution for Eq. (\ref{BSfromFss})
is given by
\begin{eqnarray}
\omega_{\textrm{BS},2}(0)=  x+y  - \frac{(\alpha +
\beta )^2}{(1-\Delta)}+ \beta   .\label{2ndBSk=0}
\end{eqnarray}
We find that when $\alpha < (1-3 \Delta)/4$, both
$\omega_{\textrm{BS},1}(0)$ and
$\omega_{\textrm{BS},2}(0)$ exist, while only
$\omega_{\textrm{BS},1}(0)$ survives when
$(1+\Delta)/4>\alpha>(1-3\Delta)/4$. Finally, for $\alpha>(1+\Delta)/4$
no bound states exist.

When $\Delta<1$ and $Q=\pi$, one finds that $F_{sc}(\pi,\omega)$=0.
Analogously, Eq. (\ref{bs}) has two solutions,
$\omega_{\textrm{BS},1}(\pi)$ and $\omega_{\textrm{BS},2}(\pi)$ when
$-(3+\Delta)/4<y-\Delta x<(1-\Delta)/4$, with explicit expressions for
their energies:
\begin{eqnarray}
\!\!\!\!\!\omega_{\textrm{BS},1}(\pi)&=& x+y+\frac{1}{2}-\frac{1+\Delta}{4}
\nonumber \\
&-&\frac{[\left(y+\frac14\right)-\Delta\left(x+\frac14\right)]^2}{1+\Delta}, \\
\label{1stBSk=pi}
\!\!\!\!\!\omega_{\textrm{BS},2}(\pi)&=&x+y+\frac{1}{2}     \nonumber \\
&+&\frac{\zeta-(2\gamma-1-\Delta)^{\frac{3}{2}}
\sqrt{2\gamma-9+9\Delta}}{8(\Delta-1)},
\label{2ndBSk=pi}
\end{eqnarray}
with
\begin{eqnarray}
\gamma&=&(y+1/4)-\Delta(x+1/4), \nonumber \\
\zeta&=&3+4\gamma-4\gamma^2-6\Delta-4\gamma\Delta+ 3\Delta^2.
\end{eqnarray}
When $y-\Delta x< -(3+\Delta)/4$ and
$(1-\Delta)/4<y-\Delta x<(3\Delta+1)/4$, only one
$\omega_{\textrm{BS},1}(\pi)$ solution exists. In case of
$y-\Delta x=(1-\Delta)/4$, $\omega_\textrm{BS,2} (\pi)$ merges with
lower boundary of the continuum.

\subsection{Propagating spin-orbital exciton states}
\label{sec:bos}

Especially at the SU(4) symmetric point, i.e., at $x=y=1/4$, spinon and 
orbiton are strongly coupled to form a joint SOEX state inside the 
spin-orbital continuum across the whole Brillouin zone. Usually such an 
elementary spin-orbital excitation in the continuum is unstable and 
decays into a spinon and an orbiton \cite{Sch12}. However, it is 
surprising that such a SOEX state propagates here as a undamped on-site 
spin-orbital excitation \cite{Herzog}, e.g. 
$S_l^-T_l^-\vert 0\rangle$ at site $l$, within the spin-orbital 
continuum with $\xi=0$ (see  Eq. (\ref{alq})). It is straightforward to 
derive,
\begin{equation}
[H,S_l^-T_l^-]\vert 0\rangle=\left[C_l(x,y)+D_l(x,y)\right]\vert 0\rangle,
\end{equation}
where
\begin{eqnarray}
C_l (x,y) &=& (x+y) S_l^- T_{l}^-  -\frac{\Delta}{4} \left(S_{l-1}^-
T_{l-1}^- + S_{l+1}^- T_{l+1}^- \right), \nonumber \\
D_l (x,y) &=&  -\frac{1}{2}\left[\Delta \left(x-\frac{1}{4}\right)
\left(S_l^-T_{l-1}^- + S_l^- T_{l+1}^-\right) \right. \nonumber \\
&+& \left. \left(y-\frac{1}{4}\right)
\left(S_{l-1}^-T_{l}^-+S_{l+1}^-T_{l}^-\right)\right].
\end{eqnarray}
The dissipative term $D_l(x,y)$ vanishes when $x=y=1/4$. Consequently,
the dispersion of the SOEX state is given by
\begin{equation}
\omega_{\textrm{SOEX}}(Q) =\frac{1}{2}\left(1-\Delta\cos Q\right).
\end{equation}

\begin{figure}[t!]
\includegraphics[width=8.2cm]{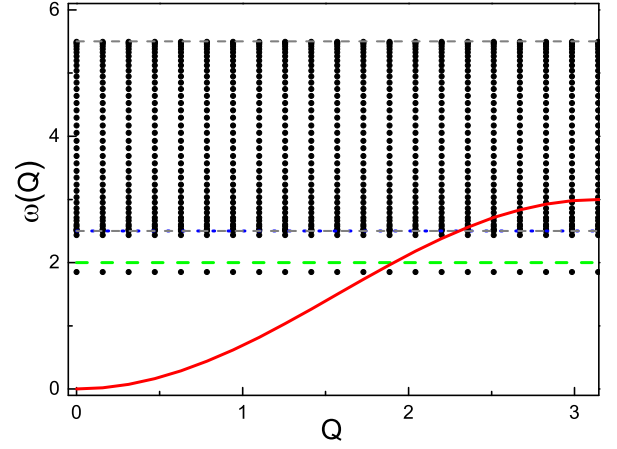}
\caption{(Color online) Excitation spectrum for the spin-orbital 
SU(2)$\otimes\mathbb{Z}_2$ model (\ref{som}) on a ring of $L=40$ sites. 
The dotted (blue) and dashed-dotted (gray) lines indicate the orbital 
excitation $\omega_T(Q)$ and the lower boundary of the continuum 
$\Omega(Q,q)$, $\omega_c(Q)$ --- both are degenerate. The dashed (green) 
and solid (red) lines denote the SOEX excitations 
$\omega_\textrm{SOEX}(Q)$, and spin excitations $\omega_S(Q)$. 
Parameters: $\Delta=0.0$, $x=0.375$, and $y=0.125$. }
\label{fig:sectr0}
\end{figure}

The SOEX state for $x=1/4$ is degenerate with the orbital wave 
excitation, see Eq. (\ref{orbex}) and Fig. \ref{fig:spectra}(a). 
When $\Delta=1$ and $x=y>0.25$, there is a quasi-SOEX state inside the 
spin-orbital continuum for $Q<\pi$, with $\xi<1$. When $x\neq y$, the 
residual signal of the SOEX state denoted by finite $\xi$ vanishes and 
the BS $\omega_\textrm{BS,2}(\pi)$ appears. The smaller the $\Delta$ is,
the more values of $Q$ give $\omega_\textrm{BS,2}(Q)$. There is no BS
at $Q=\pi$ for $y-\Delta x>(3\Delta+1)/4$. Furthermore, away from the
symmetric point, the SOEX will acquire a finite linewidth due to 
residual interactions into magnon-orbiton pairs,
\begin{equation}
\Gamma=\Im \{G^{-1}(Q,\omega)\}.
\end{equation}
The exciton spectral weight can be calculated from the self-energy 
$\Sigma(Q,\omega)$,
\begin{equation}
z_Q=\left.\left[1-\left(\frac{\partial\Sigma(Q,\omega)}
{\partial\omega}\right)\right]^{-1}\right\vert_{\omega_{\textrm{SOEX}}(Q)},
\end{equation}
where the SOEX energy is given by by the pole,
\begin{equation}
\omega_{\textrm{SOEX}}(Q)=\Re \{G^{-1}(Q,\omega)\}.
\end{equation}
If $\Gamma/[\omega-\omega_{\textrm{SOEX}}(Q)]^2 \to 0$, the exciton is 
stable. The decay rate of the SOEX increases with growing $x>1/4$ and
also for decreasing momenta $Q$, and they coincide at $\Delta=0$.

In the limit of Ising orbital interactions ($\Delta=0$), the orbital 
part becomes classical and orbitons are dispersionless, indicating 
localized orbital excitations, see Fig. \ref{fig:sectr0}. We find that 
two BSs, $\omega_{\textrm{BS},1}(0)$ and $\omega_{\textrm{BS},2}(0)$, 
exist when $y<1/4$, and there are no BSs otherwise. In contrast, at 
$Q=\pi$ one finds two solutions, $\omega_{\textrm{BS},1}(\pi)$ and 
$\omega_{\textrm{BS},2}(\pi)$, when $-1/4<y<1/4$ and no BS is found for 
$y>1/4$. In Fig. \ref{fig:sectr0}, both BSs, $\omega_{\textrm{BS,1}}$ 
and $\omega_{\textrm{BS,2}}$, are undamped. In this case, 
$\omega_T(Q)=1/4+x$ and is degenerate with the lower boundary of the 
continuum $\omega_c(Q)=1/4+x$. Moreover, 
$\omega_{\textrm{SOEX}}(Q)=x+y$, and especially when $x=y=1/4$, these 
excitations coincide. In this case all excitations are dispersionless, 
see Fig. \ref{fig:sectr0}. The spin-orbital BS in Fig. \ref{fig:sectr0} 
appears below the bottom
of the spin-orbital continuum and is stabilized by its binding energy.

\begin{figure}[t!]
\begin{center}
\includegraphics[width=8.2cm]{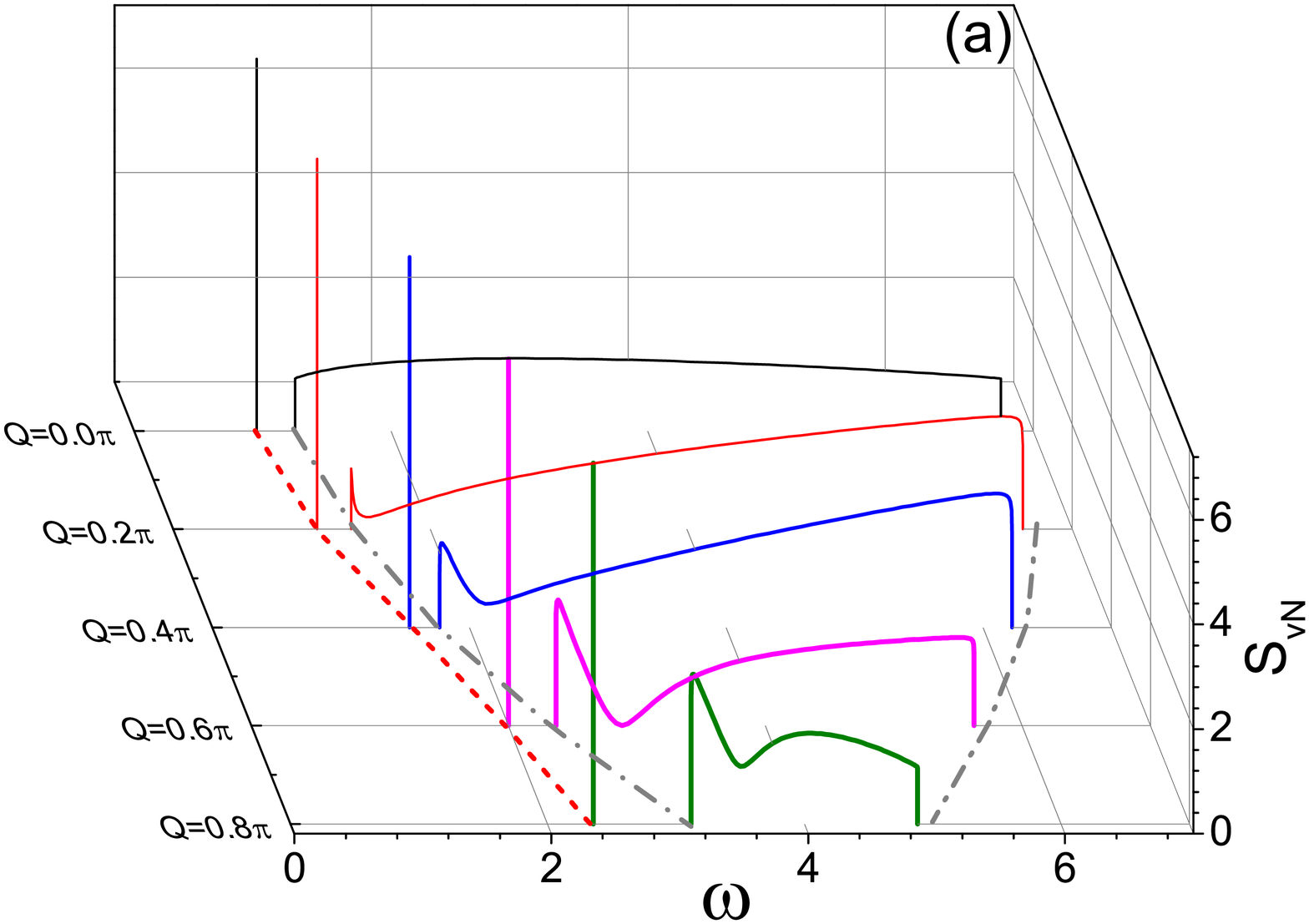}
\includegraphics[width=8.2cm]{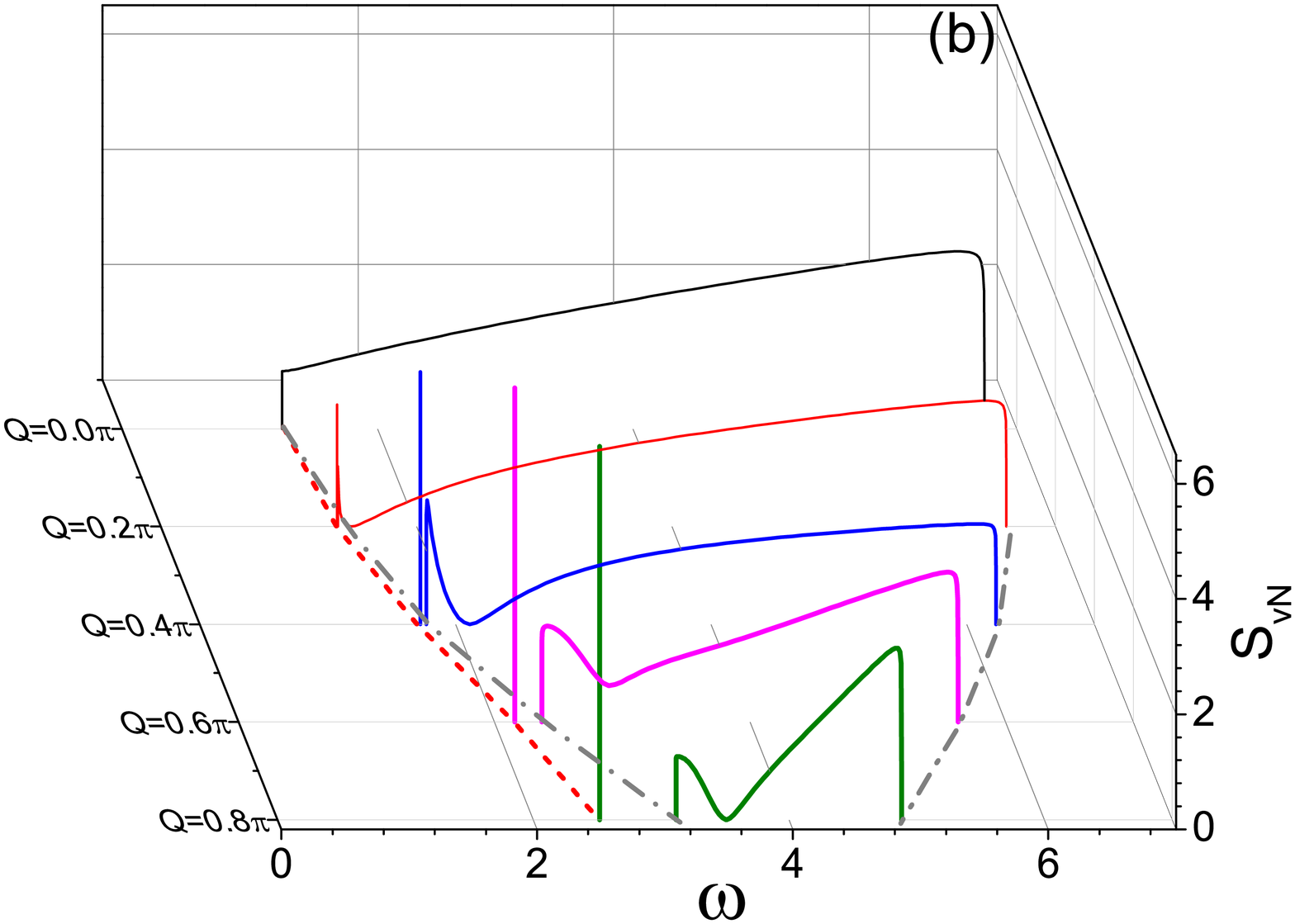}
\caption{(Color online)
The vNE spectral function ${\cal S}_{\rm vN}(Q,\omega)$ (\ref{spectra})
for the anisotropic spin-orbital SU(2)$\otimes XXZ$ model (\ref{som}) 
with $\Delta=0.5$ on a ring of $L=160$ sites, see Fig.
\ref{fig:spectra}(b), obtained for different momenta $Q\le 0.8\pi$, 
and for:
(a) even excitations, and
(b) odd excitations.
Isolated vertical lines below the continuum indicate the BS, with
dispersion given by the dashed (red) line.
Parameters: $x=0.375$, $y=0.125$.}
\label{fig:soe}
\end{center}
\end{figure}

\section{von Neumann entropy spectra}
\label{sec:entropy}

\subsection{The spectra in the FM/FO phase}
\label{sec:ffen}

To investigate the degree of entanglement of spin-orbital excited states,
we introduce the vNE spectral function in the Lehmann representation,
\begin{equation}
{\cal S}_{\rm vN}(Q,\omega)=-\sum_n \textrm{Tr}\{\rho_S^{(\mu)}
\log_2\rho_S^{(\mu)}\}\delta\left\{\omega-\omega_n(Q)\right\},
\label{spectra}
\end{equation}
where we use a short-hand notation $(\mu)=(Q,\omega_n)$ for momentum 
$Q$ and excitation energy $\omega$, and 
\begin{equation}
\rho^{(n)}_S=\textrm{Tr}_T|\Psi_n(Q)\rangle\langle\Psi_n(Q)| 
\label{rhon}
\end{equation}
is the spin ($S$) density matrix obtained by tracing over the orbital 
($T$) degrees of freedom. In Fig. \ref{fig:soe} we present the analytic 
results for the vNE spectral function when $\Delta=0.5$, $x=0.375$, and 
$y=0.125$. The parity symmetry is broken at $x\neq y$ and 
$\Delta\neq 1$. The even and odd excitations show diverse behavior of 
their entanglement, as is displayed in Fig. \ref{fig:soe}. Inspection 
of the vNE spectra shows that the entanglement reaches a local maximum 
at the BSs and SOEX states, and all these states have short-range 
correlation length $\xi$.

\begin{figure}[t!]
\includegraphics[width=8.2cm]{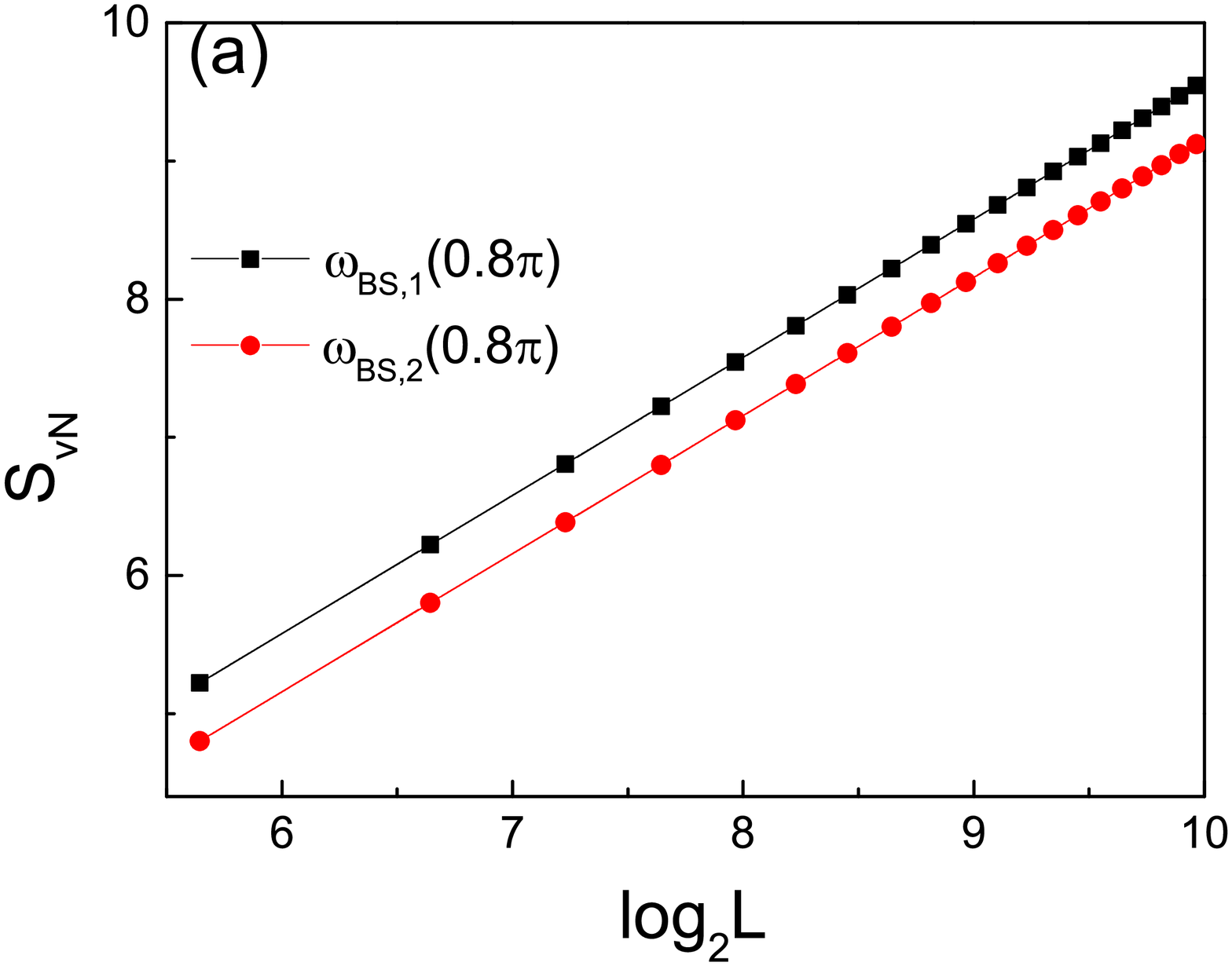}
\includegraphics[width=8.2cm]{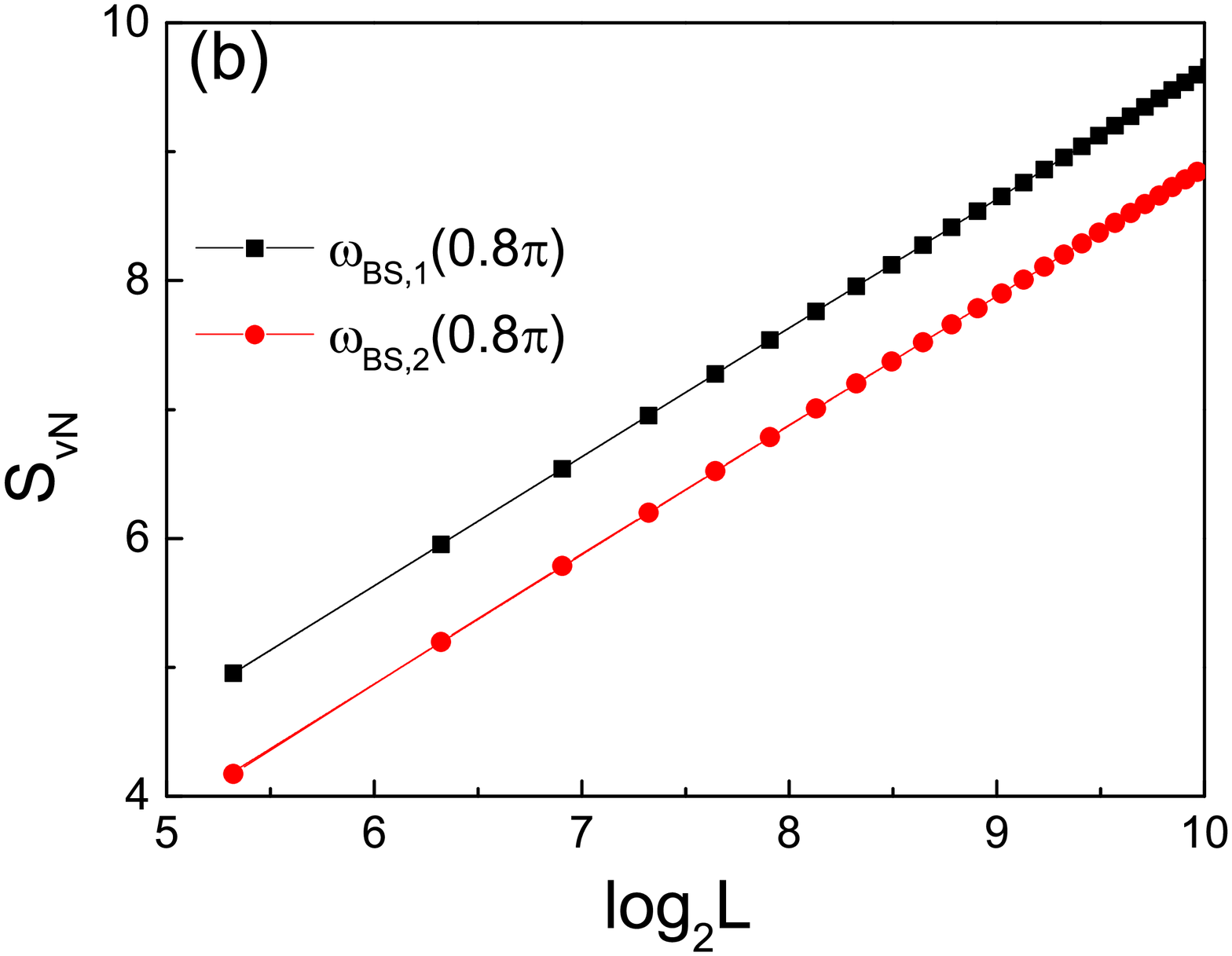}
\caption{(Color online) Scaling behavior of the entanglement entropy
${\cal S}_{\rm vN}$ of the spin-orbital BSs at $Q=0.8\pi$ (points) for
$x=0.375$, $y=0.125$, and for:
(a) $\Delta=0.5$ and 
(b) $\Delta=0$.
Lines represent logarithmic fits to Eq. (\ref{scaling}) with:
(a) $c_0=-0.421$ and $-0.842$;
(b)~$c_0=-0.368$ and $-1.122$.
}
\label{fig:scale}
\end{figure}

We have derived an asymptotic form of the vNE
as a function of $\xi$ \cite{You12},
\begin{equation}
{\cal S}_{\textrm{vN}}\simeq \log_2 \left\{\frac{L}{(1+\xi)}\right\}.
\end{equation}
One finds the asymptotic logarithmic scaling of vNE of spin-orbital BSs 
and SOEX state whose magnon-orbitons correlation length is short-range, 
and the vNE is given by
\begin{equation}
\label{scaling}
{\cal S}_\textrm{vN}= \log_2 L + c_0.
\end{equation}
In particular, $c_0=0$ for the SOEX state and $c_0<0$ otherwise.
Such relation is displayed in Fig. \ref{fig:scale}. Fig. \ref{fig:xi}
implies that  $\omega_{\textrm{BS,1}}$ are always stable for all
momenta while $\omega_{\textrm{BS,2}}$ are undamped for large momenta.

When both $x$ and $y$ are away from $1/4$, the SOEX state is unstable
and decays into a spinon and an orbiton. In this case the correlation 
length $\xi$ becomes extensive. We have verified that the scaling is 
entirely different in such a case and the entropy of the SOEX scales 
instead as a power law,
\begin{equation}
{\cal S}_\textrm{vN}=\frac{c_1}{L}+c_0.
\end{equation}
The vNE saturates in the thermodynamic limit.

\subsection{RIXS spectral functions in the FM/FO state}
\label{sec:rixs}

The entanglement spectral function ${\cal S}_{\rm vN}(Q,\omega)$ has a
similar form as any other dynamical spin or charge correlation function.
There is, however, an important difference --- as there is no direct
probe for the vNE of an arbitrary state, the SOE spectra can be
calculated but cannot be measured directly. On the other hand, we have
shown before \cite{You12} that the intensity distribution of certain
RIXS spectra of spin-orbital excitations in fact probe qualitatively
SOE.

\begin{figure}[b!]
\begin{center}
\includegraphics[width=8.4cm]{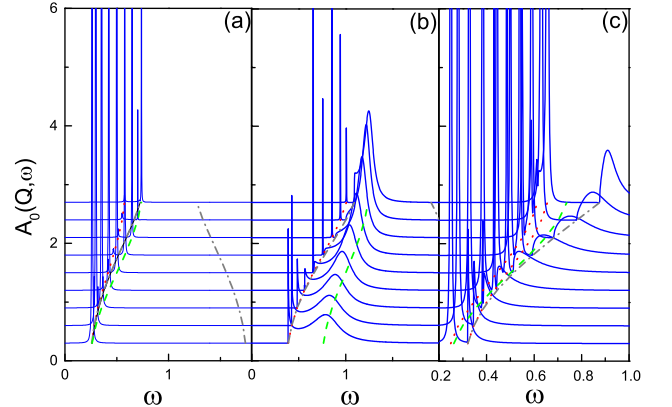}
\caption{(Color online) The spectral function of the on-site
spin-orbital excitation $A_0(Q,\omega)$ for the anisotropic
SU(2)$\otimes XXZ$ model with $\Delta=0.5$, and for:
(a) $x=0.25$, $y=0.25$,
(b) $x=0.5$, $y=0.5$, and
(c) $x=0.375$, $y=0.125$.
In each panel the momenta $Q$ range from $\pi/10$ (bottom) to $9\pi/10$
(top); the peak broadening is $\eta=0.001$. The red dotted (green
dashed) lines mark the positions of BSs (SOEX) states. Gray dash-dot
line signals the onset of the continuum.}
\label{fig:a0}
\end{center}
\end{figure}

We introduce the spectral function of the coupled spin-orbital 
excitations at distance $l$,
\begin{equation}
A_l(Q,\omega) =\frac{1}{\pi} \lim_{\eta \to 0} \textrm{Im}
\langle 0\vert\Gamma_Q^{(l)\dagger}\frac{1}{\omega+E_0 -H -i \eta}\,
\Gamma_Q^{(l)} \vert 0\rangle.
\end{equation}
Here
\begin{equation}
\Gamma_Q^{(0)}=\frac{1}{\sqrt{L}}\sum_j e^{iQj} S_j^- T_j^-
\end{equation}
is the local operator for an on-site joint spin-orbital excitation 
measured in RIXS \cite{Ame09,Bis15,Che15}. We employ as well the even 
and odd parity operators,
\begin{equation}
\Gamma_Q^{(1{\pm})}=\frac{1}{\sqrt{2L}}
\sum_j e^{iQj}\left(S_{j+1}^-\pm S_{j-1}^-\right) T_j^-,
\end{equation}
which serve to probe the nearest neighbor spin-orbital excitations.

\begin{figure}[t!]
\begin{center}
\includegraphics[width=8.4cm]{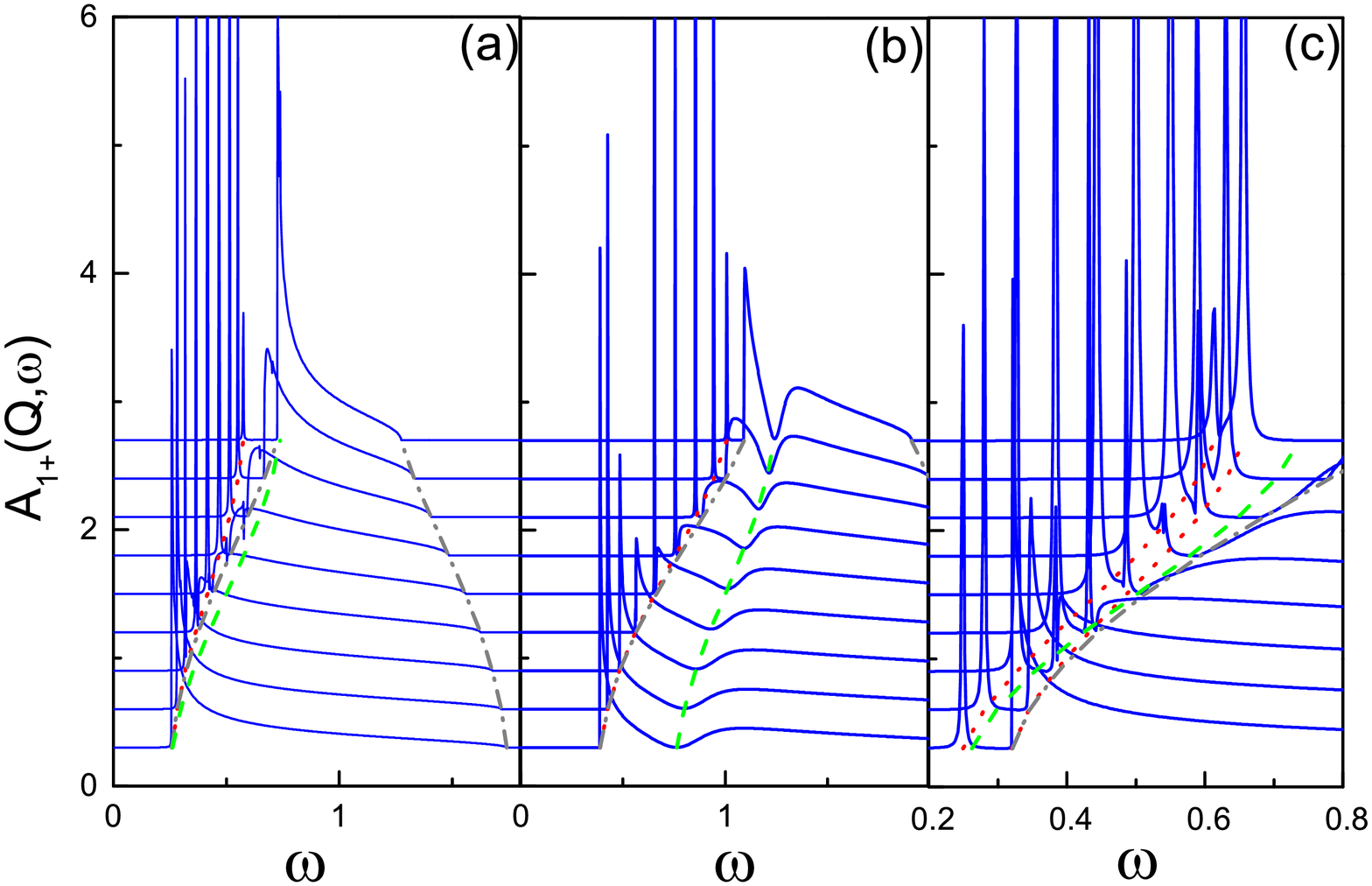}
\caption{(Color online) The spectral function of the spin-orbital
excitation at nearest neighbors $A_{1+}(Q,\omega)$ with even parity for
the anisotropic SU(2)$\otimes XXZ$ model with $\Delta=0.5$, and for:
(a) $x=0.25$, $y=0.25$,
(b) $x=0.5$, $y=0.5$, and
(c) $x=0.375$, $y=0.125$.
In each panel the momenta $Q$ range from $\pi/10$ (bottom) to $9\pi/10$
(top); the peak broadening is $\eta=0.001$. The red dotted (green
dashed) lines mark the positions of BSs (SOEX) states. Gray dash-dot
line signals the onset of the continuum. }
\label{fig:a1+}
\end{center}
\end{figure}

\begin{figure}[b!]
\begin{center}
\includegraphics[width=8.4cm]{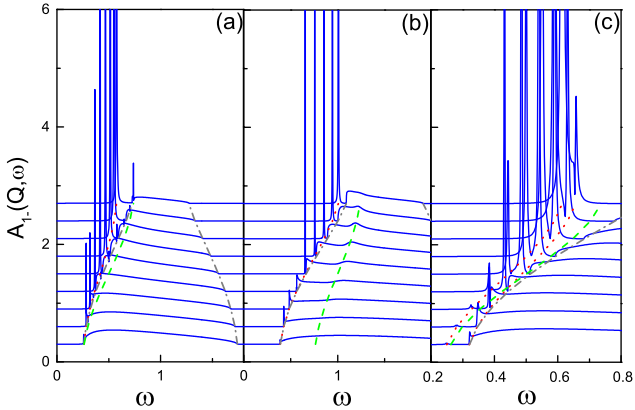}
\caption{(Color online) The spectral function of the spin-orbital
excitation at nearest neighbors $A_{1+}(Q,\omega)$ with odd parity for
the anisotropic SU(2)$\otimes XXZ$ model with $\Delta=0.5$, and for:
(a) $x=0.25$, $y=0.25$,
(b) $x=0.5$, $y=0.5$, and
(c) $x=0.375$, $y=0.125$.
In each panel the momenta $Q$ range from $\pi/10$ (bottom) to $9\pi/10$
(top); the peak broadening is $\eta=0.001$. The red dotted (green
dashed) lines mark the positions of BSs (SOEX) states. Gray dash-dot
line signals the onset of the continuum.  }
\label{fig:a1-}
\end{center}
\end{figure}

Intuitively, the on-site spectral function $A_{0}(Q,\omega)$ highlights
the SOEX state. It is found that both $\omega_{\textrm{SOEX}}(Q)$ and
$\omega_{\textrm{BS}}(Q)$ are solutions of Eq. (\ref{bs}) when 
$x=y=1/4$. However, the weight of the BS is here zero. Now the spectral
function is given by
\begin{eqnarray}
A_0(Q,\omega)=\delta\left\{\omega-\omega_{\textrm{SOEX}}(Q)\right\},
\end{eqnarray}
which is confirmed in Fig. \ref{fig:a0}(a). As the point $(x,y)$ moves 
away from the symmetric point, i.e., $x=y=1/4$, the BSs gain spectral 
weight which decreases with momentum $Q$. The spectral weight vanishes 
at $Q=\pi$ accompanying a square-root singularity at the lower bound of
the continuum, see Fig. \ref{fig:a0}(b). The $\delta$ peak turns into a
broad peak. A difference between $x$ and $y$ induces a second branch of
BSs and it gains larger spectral weight at large momenta than the first 
BS, see Fig. \ref{fig:a0}(c). Altogether, the evolution of spectral 
weight is similar to that of the correlation length in Fig. \ref{fig:xi}.

The BSs can be captured also by the spectral functions for the nearest
neighbor excitations, see Figs. \ref{fig:a1+} and \ref{fig:a1-}. With
increasing momentum $Q$, the spectral weight of even-parity excitation
$A_{1+}(Q,\omega)$ at the first BS decreases (see Fig. \ref{fig:a1+}),
while the spectral weight of odd-parity excitation $A_{1-}(Q,\omega)$
rises (see Fig. \ref{fig:a1-}). $A_{1+}(Q,\omega)$ reaches its valley
for quasi-SOEX states, but no such feature could be found in
$A_{1-}(Q,\omega)$.

\section{Discussion and conclusions}
\label{sec:summa}

Motivated by the discovery of new Majumdar-Ghosh-like valence-bond spin 
singlet phases triggered by orbital correlations, we have studied the
spin-orbital entanglement (SOE) in the one-dimensional (1D)
anisotropic SU(2)$\otimes XXZ$ spin-orbital model with the negative
exchange interaction. The asymmetry between spin and orbital degrees
of freedom yields a better insight into the phase diagram and the
mechanisms responsible for the different types of order observed for
this system. In addition to the four uniform phases I-IV, our study 
demonstrates that a gapful phase V exists in case of classical Ising 
orbital interactions, i.e.,  in the SU(2)$\otimes\mathbb{Z}_2$ model. 
It is characterized by quadrupling of the unit cell seen as a maximum 
of the orbital structure factor at $k=\pi/2$. For $y=-1/4$ this 
provides a perfect dimer structure of spin singlets in the whole region 
of stability of this phase, where the dimer spin correlations $D(r)$ 
develop and uncover long-range dimer order. The dimer phase V is quite 
robust and survives when the orbital quantum fluctuations at
$\Delta>0$ are taken into account.

The phase diagram is still richer at finite $\Delta>0$, when quantum 
orbital fluctuations develop and induce an orbital dimer phase VI, with 
a complementary role of spin and orbital correlations to phase V. 
The emergence of the nonuniform phase V is a result of the joint 
interaction between spin fluctuation and orbital degree of freedom, and 
thus phase V carries finite SOE. The orbital fluctuations enhance the 
SOE in ground state near the III/V phase transition and lead to phase 
VI when the $\{x,y\}$ parameters are interchanged. We also anticipate 
that these dimer phases may survive in higher dimensions. In fact, 
Lieb-Schultz-Mattis theorem is applicable to the present model for
arbitrary $x$ and $y$ and nonzero $\Delta$, where a finite gap exists
above these degenerate ground states. Both phases V and VI are gapped
phases with alternating spin and orbital singlets, respectively.
As we have shown, the phase boundaries can be captured by SOE and
fidelity susceptibility. The phase transition between phases III and V
is a first-order transition in the $\Delta=0$ case, and the transition
changes to continuous when $\Delta>0$.

An important consequence of finite SOE in the ground state is that it
invalidates mean-field decoupling of spin and orbital degrees of
freedom, as this would imply a spin-orbital product ground state.
A similar restriction applies to the entangled elementary excitations 
in the disentangled ferromagnetic phase with ferro-orbital order in 
spin-orbital systems which were analyzed here with the help of the von 
Neumann entropy spectral function. Spin-orbital excitations are 
highlighted by nontrivial SOE, especially by logarithmic scaling of 
SOE in this phase.

\textit{A priori}, the SOE makes it necessary to treat the eigenstates
of a given model exactly. In fact, since a mean-field decoupling
shields the SOE, it fails to describe elementary excitations even
qualitatively correctly in a number of spin-orbital models.
In antiferromagnetic ground states with ferro-orbital order it was 
demonstrated both in theory \cite{Woh11} and experiment \cite{Sch12} 
that the spin-orbital excitation fractionalizes into freely propagating 
spinon and orbiton, giving rise to spin-orbital separation under 
specific condition. The SOE in the spin-orbital separation remains 
unclear. The low-lying excitations in phases II (AF/FO) and III (AF/AO) 
are spin waves with vanishing SOE, corresponding to a two-spinon 
continuum of an antiferromagnetic spin chain. The low-lying excitation 
in phases V and VI corresponds to spin-orbital excitation, as shown in 
Figs. \ref{fig:scheme}(b) and \ref{fig:scheme}(c). 
The problem of the SOE in elementary excitations in other phases 
remains a challenge for future studies.

Summarizing, we have shown that the anisotropic SU(2)$\otimes XXZ$
spin-orbital model with \textit{negative} exchange coupling has
remarkably different behavior and phase diagrams from the well known 
SU(2)$\otimes$SU(2) model with \textit{positive} exchange coupling.
While the spin-orbital liquid phase is absent in the former case, we
have found that the joint ferromagnetic/ferro-orbital fluctuations are 
surprisingly strong at the quantum phase transition to the 
antiferromagnetic spin order which gives even stronger SOE than that 
established for the 1D isotropic SU(2)$\otimes$SU(2)model with positive 
exchange coupling.

\acknowledgments

We thank Krzysztof Wohlfeld for insightful discussions.
W.-L.Y. acknowledges support by the Natural Science Foundation
of Jiangsu Province of China under Grant No.~BK20141190 and the
NSFC under Grant No.~11474211. A.M.O. kindly acknowledges
support by Narodowe Centrum Nauki (NCN, National Science Center)
under Project No. 2012/04/A/ST3/00331.

\appendix*

\section{Phase diagrams for the two-site model}
\label{sec:appa}

\begin{figure}[t!]
\begin{center}
\includegraphics[width=7.2cm]{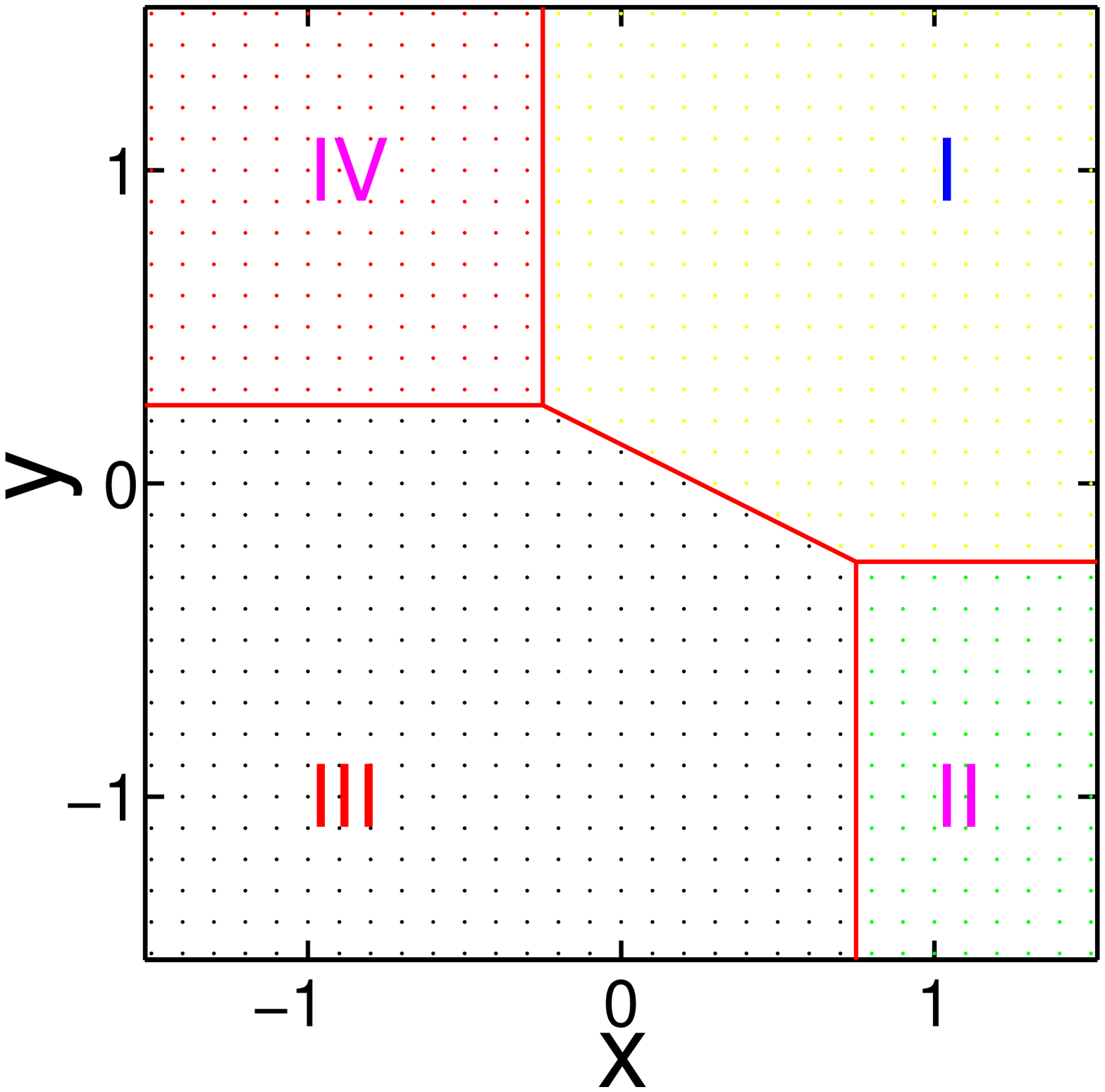}
\includegraphics[width=7.2cm]{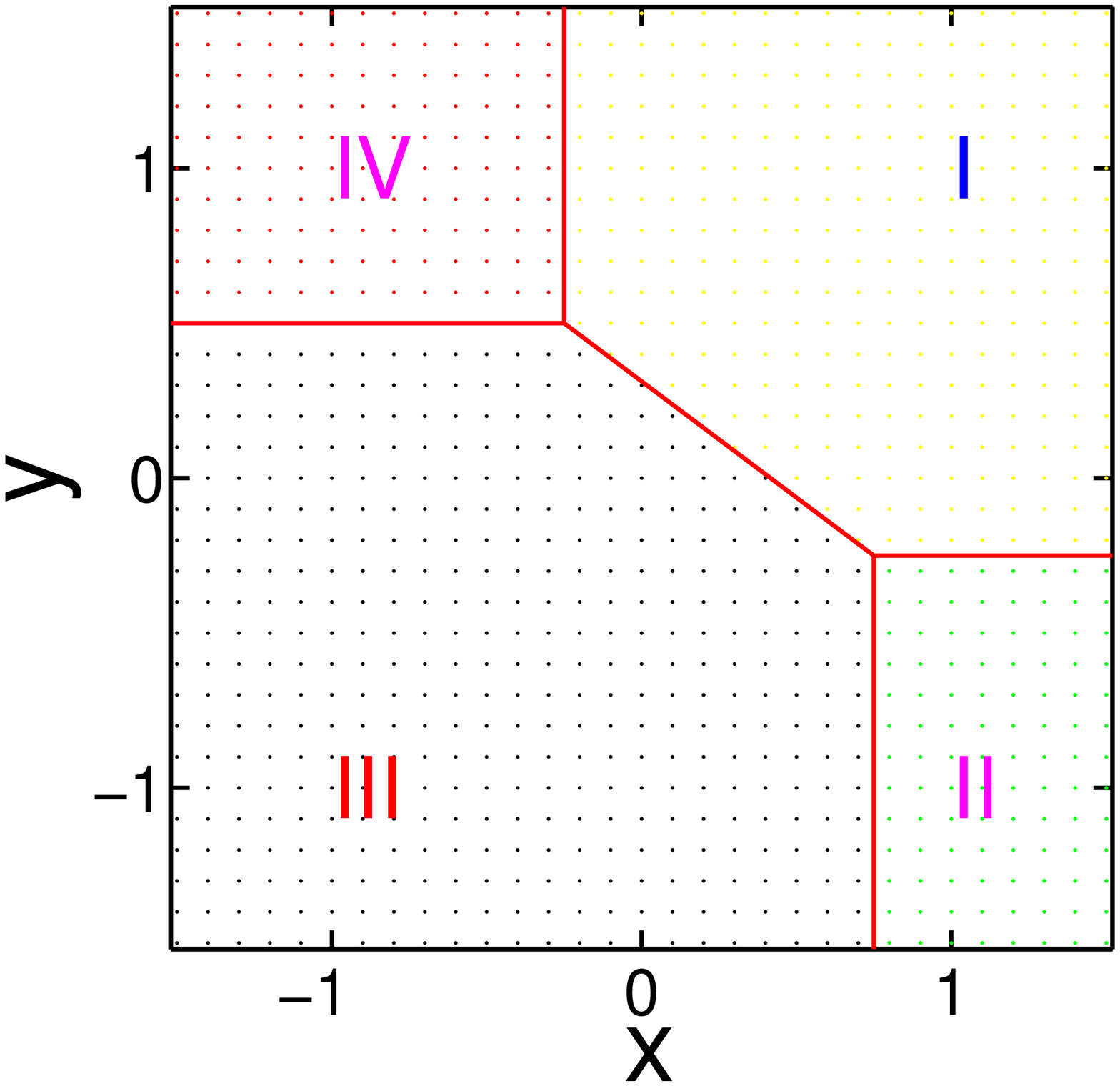}
\caption{(Color online)(a) The phase diagram of 2-site model
when $\Delta=0$.
(b) $\Delta=0.5$. Phases I-IV correspond to FS/FO,AS/FO,
AS/AO, FS/AO configurations in spin-orbital sectors. }
\label{fig:2site}
\end{center}
\end{figure}

To understand better the phase boundaries in the anisotropic
SU(2)$\otimes XXZ$ spin-orbital model with negative exchange coupling
(\ref{som}), we present here an exact solution for the system of $L=2$
sites \cite{Brink},
\begin{eqnarray}
\label{som12}
{\cal H}_{12}&=\!&- J\left(\vec{S}_{1}\cdot\vec{S}_{2}+x\right) \nonumber \\
&\times\!&\left[\frac{\Delta}{2}
\left(T_1^+T_2^-+T_1^-T_2^+\right)+T_1^zT_2^z+y\right],
\end{eqnarray}
Again, $\{x,y\}$ are the parameters, and $0\leq\Delta\leq 1$
interpolates between the Ising $\mathbb{Z}_2$ ($\Delta=0$) and 
Heisenberg SU(2) ($\Delta=1$) symmetry. The orbital interaction with 
$XXZ$ symmetry can be exactly diagonalized and one finds 4 eigenstates:
$\vert + + \rangle$,
$\vert - - \rangle$,
$ \left(\vert + -\rangle + \vert  - +\rangle\right)/\sqrt{2}$,
$ \left(\vert + -\rangle - \vert  - +\rangle\right)/\sqrt{2}$,
corresponding to eigenvalues:
$1/4$, $1/4$, $\Delta/2-1/4$, $-\Delta/2-1/4$, respectively. At 
$\Delta=0$ we recover doubly degenerate configurations from the latter 
two, while at $\Delta=1$ we recover a triplet $T=1$ from the first 
three. In any case, the third component of the triplet,
$ \left(\vert + -\rangle + \vert  - +\rangle \right)/\sqrt{2}$,
is always an entangled excited state with the present choice of
parameters, while the orbital singlet,
$ \left(\vert + -\rangle - \vert  - +\rangle\right)/\sqrt{2}$,
is an entangled ground state for some parameters.

The spin part is classified as a triplet $S=1$ or a singlet $S=0$, and
these states are accompanied by the orbital states described above.
This gives the states I-IV in Table II, and one of them is the ground
state for any point in the $(x,y)$ plane. All phase transitions are
first order, with a change of spin or orbital state. The phases II and
IV are symmetric for $\Delta=1$, and the transition between I and III
occurs along the $x+y=1/2$ line \cite{You12}. At $\Delta=0$ phase IV
exists for $y>1/4$ while phase II only for $x>3/4$,
see Fig. \ref{fig:2site}(a). This reflects the essential difference
between the orbital configurations in the Ising limit and the quantum
spin states. Note that spin singlet $S=0$ is an entangled state, but
the larger Hilbert space gives no spin-orbital entanglement in any
phase. Thus, the total energies and the phase diagram are easily
deduced, see Fig. \ref{fig:2site}.

\begin{table}[t!]
\caption{The spin-orbital configuration for the model (\ref{som12}),
with phases I-IV defined by distinct spin $S$ and orbital state as
as obtained at $0<\Delta<1$. The lowest energy $E^0/J$ has degeneracy
$d$ and becomes the ground state at the respective values of $\{x,y\}$
parameters, Fig. \ref{fig:2site}. At $\Delta=0$ the degeneracies of
phases III and IV change to 2 and 6 for the
$\mathbb{Z}_2\otimes\mathbb{Z}_2$ symmetry, while at $\Delta=1$ the
degeneracies for the ground states I-IV are 9, 3, 1, 3 and follow from
the SU(2)$\otimes$SU(2) symmetry.
}
\label{tab:2site}
\begin{ruledtabular}
\begin{tabular}{ccccccc}
& phase & S & orbital states & $E^0/J$ & $d$ & \\
\hline
&  I  & 1 & $\vert+ +\rangle$, $\vert- -\rangle$ &
 $-\left( \frac14+x \right)\left( \frac14+y \right)$ & 6 &\cr
&  II & 0 & $\vert+ +\rangle$, $\vert- -\rangle$ &
 $-\left(-\frac34+x \right)\left( \frac14+y \right)$ & 2 &\cr
& III & 0 & $\frac{1}{\sqrt{2}}\left(\vert + -\rangle-\vert - +\rangle\right)$ &
 $ \left(-\frac34+x \right)\left( \frac14+\frac{\Delta}{2}-y \right)$ & 1 &\cr
&  IV & 1 & $\frac{1}{\sqrt{2}}\left(\vert + -\rangle-\vert - +\rangle\right)$ &
 $ \left( \frac14+x \right)\left( \frac14+\frac{\Delta}{2}-y \right)$ & 3 &\cr
\end{tabular}
\end{ruledtabular}
\end{table}

From the comparison of the energies $E^0$ of phases III and IV (Table
II) one can see that the boundary between III and IV, given by the
straight line $y=1/4+\Delta/2$, moves upwards with increasing $\Delta$,
see Fig. \ref{fig:2site}(b). Accordingly, the phase boundary between
phases I and III is also modified. The interplay between spins and
orbitals develops and leads to interesting consequences of entanglement
for rings with $L\ge 4$.
Remarkably, the trivial phase diagram found for Ising-Ising interactions
for $L=2$ sites (with II-III and IV-III transition lines at $x=1/4$ and
$y=1/4$, similar to the diagrams in Fig. \ref{fig:2site}) is the same
for the $\mathbb{Z}_2\otimes\mathbb{Z}_2$ spin-orbital model in the
thermodynamic limit,
\begin{equation}
{\cal H}_{\mathbb{Z}_2\otimes\mathbb{Z}_2}=
-J\sum_{j}\left(S_j^z S_{j+1}^z+x\right)\left(T_j^z T_{j+1}^z+y\right).
\label{II-Ham}
\end{equation}


\begin{thebibliography}{99}

\bibitem{Kug82} K. I. Kugel and D. I. Khomskii,
                   JETP \textbf{37}, 725 (1973);
                   Sov. Phys. Usp. \textbf{25}, 231 (1982).

\bibitem{Fei97} L. F. Feiner, A. M. Ole\'s, and J. Zaanen,
                   Phys. Rev. Lett. \textbf{78}, 2799 (1997);
                   J. Phys.: Condens. Matter \textbf{10}, L555 (1998).

\bibitem{Tokura} Y. Tokura and N. Nagaosa,
                   Science \textbf{288}, 462 (2000).

\bibitem{Ole05} A. M. Ole\'s, G. Khaliullin, P. Horsch, and L. F. Feiner,
                   Phys. Rev. B \textbf{72}, 214431 (2005).

\bibitem{Kha05} G. Khaliullin,
                   Prog. Theor. Phys. Suppl. \textbf{160}, 155 (2005).

\bibitem{Ole12} A. M. Ole\'s,
                   J. Phys.: Condens. Matter \textbf{24}, 313201 (2012);
                   Acta~Phys. Polon. A  \textbf{127}, 163 (2015).

\bibitem{Li14}  Yi Li, E. H. Lieb, and C. Wu,
                   Phys. Rev. Lett. \textbf{112}, 217201 (2014).

\bibitem{Zhao}  E. Zhao and W. V. Liu,
                   Phys. Rev. Lett. \textbf{100}, 160403 (2008).

\bibitem{Wu08}  C. Wu,
                   Phys. Rev. Lett. \textbf{100}, 200406 (2008);
                C. Wu and S. Das Sarma,
                   Phys. Rev. B \textbf{77}, 235107 (2008).

\bibitem{Sun12} G. Sun, G. Jackeli, L. Santos, and T. Vekua,
                   Phys. Rev. B \textbf{86}, 155159 (2012).

\bibitem{Victor} V. Galitski and I. B. Spielman,
                   Nature \textbf{494}, 49 (2013).

\bibitem{Zho15} Z. Zhou, E. Zhao, and W. V. Liu,
                   Phys. Rev. Lett. \textbf{114}, 100406 (2015).

\bibitem{Jac09} G. Jackeli and G. Khaliullin,
                   Phys. Rev. Lett. \textbf{102}, 017205 (2009).

\bibitem{Hasan} M. Z. Hasan and C. L. Kane,
                   Rev. Mod. Phys. \textbf{82}, 3045 (2010).

\bibitem{Kim}   B. J. Kim, H. Jin, S. J. Moon, J.-Y. Kim, B.-G. Park,
                   C.~S.~Leem, J. Yu, T. W. Noh, C. Kim, S.-J. Oh,
                   J.-H.~Park, V. Durairaj, G.~Cao, and E. Rotenberg,
                   Phys. Rev. Lett. \textbf{101}, 076402 (2008).

\bibitem{Comin} R. Comin, G. Levy, B. Ludbrook, Z.-H. Zhu, C. N. Veenstra,
                   J.~A.~Rosen, Y. Singh, P. Gegenwart, D. Stricker,
                   J.~N.~Hancock, D. van der Marel, I. S. Elfimov,
                   and A. Damascelli,
                   Phys. Rev. Lett. \textbf{109}, 266406 (2012);
                F.~Trousselet, M. Berciu, A. M. Ole\'s, and P. Horsch,
                   {\it ibid.\/} \textbf{111}, 037205 (2013).

\bibitem{Kha13} G. Khaliullin,
                   Phys. Rev. Lett. \textbf{111}, 197201 (2013).

\bibitem{Akb14} A. Akbari and G. Khaliullin,
                   Phys. Rev. B \textbf{90}, 035137 (2014).

\bibitem{Ole06} A. M. Ole\'s, P. Horsch, L. F.~Feiner, and G. Khaliullin,
                   Phys. Rev. Lett. \textbf{96}, 147205 (2006).

\bibitem{Brz14} W. Brzezicki, J. Dziarmaga, and A. M. Ole\'s,
                   Phys. Rev. Lett. \textbf{112}, 117204 (2014).

\bibitem{Kag14} M. Y. Kagan, K. I. Kugel, A. V. Mikheyenkov, 
                   and A.~F.~Barabanov,
                   JETP Lett. \textbf{100}, 187 (2014).

\bibitem{Brz15} W. Brzezicki, A. M. Ole\'s, and M. Cuoco,
                   Phys. Rev. X \textbf{5}, 011037 (2015).

\bibitem{Muh09} S. M\"{u}hlbauer, B. Binz, F. Jonietz, C. Pfleiderer,
                   A.~Rosch, A.~Neubauer, R. Georgii, and P. B\"{o}ni,
                   Science \textbf{323}, 915 (2009).

\bibitem{Xue13} C. Z. Chang, J. Zhang, X. Feng, J. Shen, Z. Zhang, M.~Guo,
                   K.~Li, Y. Ou, P. Wei, L. L. Wang, Z.~Q.~Ji, Y.~Feng,
                   S. Ji, X.~Chen, J. Jia, X. Dai, Z.~Fang, S.~C.~Zhang,
                   K.~He, Y. Wang, L.~Lu, X. C. Ma, and Q.~K.~Xue,
                   Science \textbf{ 340}, 167 (2013).

\bibitem{Wan11} X. Wan, A. M. Turner, A. Vishwanath, and S.~Y.~Savrasov,
                   Phys. Rev. B \textbf{83}, 205101  (2011).

\bibitem{Zhu13} Z.-H. Zhu, C. N. Veenstra, G. Levy, A. Ubaldini,
                   P. Syers, N.~P.~Butch, J. Paglione, M. W. Haverkort,
                   I. S. Elfimov, and A.~Damascelli,
                   Phys. Rev. Lett. \textbf{110}, 216401 (2013).

\bibitem{Gas13} Z.-H. Zhu, A. Nicolaou, G. Levy, N. P. Butch, P. Syers,
                   X.~F.~Wang, J. Paglione, G. A. Sawatzky,
                   I. S. Elfimov, and A.~Damascelli,
                   Phys. Rev. Lett. \textbf{111}, 216402 (2013).

\bibitem{Yuki}  Y. Ishiguro, K. Kimura, S. Nakatsuji, S. Tsutsui,
                   A.~Q.~R.~Baron, T. Kimura, and Y. Wakabayashi,
                   Nat. Commun. \textbf{4}, 3022 (2013).

\bibitem{Kat15} N. Katayama, K. Kimura, Y. Han, J. Nasu, N. Drichko,
                   Y. Nakanishi, M. Halim, Y. Ishiguro, R. Satake,
                   E. Nishibori, M. Yoshizawa, T. Nakano, Y. Nozue,
                   Y. Wakabayashi, S. Ishihara, M. Hagiwara, H. Sawa,
                   S. Nakatsuji, 
                   Proc. Nat. Ac. Sci. \textbf{112}, in press (2015).

\bibitem{Leon}  Leon Balents,
                   Nature \textbf{464}, 199 (2010).

\bibitem{Mit14} L. Mittelst\"adt, M. Schmidt, Z. Wang, F. Mayr, V.~Tsurkan,
                   P.~Lunkenheimer, D. Ish, L. Balents, J.~Deisenhofer,
                   and A.~Loidl, Phys. Rev. B \textbf{91}, 125112 (2015).

\bibitem{Lau14} N. J. Laurita, J. Deisenhofer, L. D. Pan, C. M. Morris,
                   M.~Schmidt, M. Johnsson, V. Tsurkan, A. Loidl,
                   and N.~P.~Armitage,
                   Phys. Rev. Lett. \textbf{114}, 207201 (2015).

\bibitem{Nor08} B. Normand and A. M. Ole\'s,
                   Phys. Rev. B \textbf{78}, 094427 (2008);
                B. Normand,
                  {\it ibid.\/} \textbf{83}, 064413 (2011);
                J.~Chaloupka and A.~M.~Ole\'s,
                  {\it ibid.\/} \textbf{83}, 094406 (2011).

\bibitem{Karlo} P. Corboz, M. Lajk\'o, A. M. L\"auchli, K. Penc, and F.~Mila,
                   Phys. Rev. X \textbf{2}, 041013 (2012).

\bibitem{Mil14} A. Smerald and F. Mila,
                   Phys. Rev. B \textbf{90}, 094422 (2014).

\bibitem{vdB11} L. J. P. Ament, M. van Veenendaal, T. P. Devereaux,
                   J.~P.~Hill, and J. van den Brink,
                   Rev. Mod. Phys. \textbf{83}, 705 (2011).

\bibitem{Woh11} K. Wohlfeld, M. Daghofer, S. Nishimoto, G. Khaliullin,
                   and J. van den Brink,
                   Phys. Rev. Lett. \textbf{107}, 147201 (2011);
                K.~Wohlfeld, S. Nishimoto, M. W. Haverkort, and J. van den Brink,
                   Phys. Rev. B \textbf{88}, 195138 (2013).

\bibitem{Sch12} J. Schlappa, K. Wohlfeld, K. J. Zhou, M. Mourigal,
                   M.~W.~Haverkort, V. N. Strocov, L. Hozoi, C. Monney,
                   S. Nishimoto, S. Singh, A. Revcolevschi, J.-S. Caux,
                   L. Patthey, H.~M.~R\o{}nnow, J. van den Brink, and
                   T.~Schmitt, Nature \textbf{485}, 82 (2012).

\bibitem{Jungho} J. Kim, D. Casa, M. H. Upton, T. Gog, Y.-J. Kim,
                   J.~F.~Mitchell, M. van Veenendaal, M. Daghofer,
                   J. van den Brink, G. Khaliullin, and B. J. Kim,
                   Phys. Rev. Lett. \textbf{108}, 177003 (2012).

\bibitem{Zhe11} Z. Wang, M. Schmidt, A. G\"unther, S. Schaile, N. Pascher,
                   F.~Mayr, Y. Goncharov, D. L. Quintero-Castro,
                   A. T. M. N. Islam, B. Lake, H.-A. Krug von Nidda,
                   A. Loidl, and J.~Deisenhofer,
                   Phys. Rev. B \textbf{83}, 201102(R) (2011).

\bibitem{Goode} J. B. Goodenough,
                   \textit{Magnetism and the Chemical Bond}
                   (Wiley Interscience, New York, 1963).

\bibitem{Amico} L. Amico, R. Fazio, A. Osterloh, and V. Vedral,
                   Rev. Mod. Phys. \textbf{80}, 517 (2008).

\bibitem{You12} W.-L. You, A. M. Ole\'s, and P. Horsch,
                   Phys. Rev. B \textbf{86}, 094412 (2012).

\bibitem{Eisert} J. Eisert, M. Cramer, and M. B. Plenio,
                   Rev. Mod. Phys. \textbf{82}, 277 (2010).

\bibitem{Mas09} Llu\'{i}s Masanes,
                   Phys. Rev. A \textbf{80}, 052104 (2009).

\bibitem{Hag15} I. Hagym\'asi, J. S\'olyom, and \"O. Legeza,
                   Phys. Rev. B \textbf{92}, 035108 (2015).

\bibitem{Chen07} Y. Chen, Z. D. Wang, Y. Q. Li, and F. C. Zhang,
                   Phys. Rev. B \textbf{75}, 195113 (2007).

\bibitem{Rex}   R. Lundgren, V. Chua, and G. A. Fiete,
                   Phys. Rev. B \textbf{86}, 224422  (2012).

\bibitem{Ber13} R. Berkovits,
                   Phys. Rev. Lett. \textbf{108}, 176803 (2012);
                   Phys. Rev. B      \textbf{87}, 075141 (2013).

\bibitem{You15} W.-L. You, A. M. Ole\'s, and P. Horsch,
                   New J. Phys. \textbf{17}, in press (2015).

\bibitem{Got10} D. Gottesman and M. B. Hastings,
                   New J. Phys. \textbf{12}, 025002 (2010).

\bibitem{Cal04} P. Calabrese and J. Cardy,
                   J. Stat. Mech. P06002 (2004).

\bibitem{Hel11} R. Helling, H. Leschke, and W. Spitzer,
                   Int. Math. Res. Not. \textbf{2011}, 1451 (2011).

\bibitem{Wu04}  L.-A. Wu, M. S. Sarandy, and D. A. Lidar,
                   Phys. Rev. Lett. \textbf{93}, 250404 (2004).

\bibitem{Alb09} V. Alba, M. Fagotti, and P. Calabrese,
                   J.~Stat. Mech. P10020 (2009).

\bibitem{Li99}  Y.-Q. Li, M. Ma, D.-N. Shi, and F.-C. Zhang,
                   Phys. Rev. Lett. \textbf{81}, 3527 (1998);
                B. Frischmuth, F. Mila, and M.~Troyer,
                   \textit{ibid.} \textbf{82}, 835 (1999).

\bibitem{Ame09} L. J. P. Ament, G. Ghiringhelli, M. M. Sala, L.~Braicovich, 
                   and J. van den Brink, 
                   Phys. Rev. Lett. \textbf{103}, 117003 (2009).

\bibitem{Bis15} V. Bisogni, K. Wohlfeld, S. Nishimoto, C. Monney, 
                   J.~Trinckauf, K. Zhou, R. Kraus, K. Koepernik, C. Sekar, 
                   V. Strocov, B. B\"uchner, T. Schmitt, J. van den Brink, 
                   and J. Geck,
                   Phys. Rev. Lett. \textbf{114}, 096402 (2015).

\bibitem{Che15} Cheng-Chien Chen, M. van Veenendaal, T. P. Devereaux, 
                   and K.~Wohlfeld, 
                   Phys. Rev. B \textbf{91}, 165102 (2015).

\bibitem{Sut75} B. Sutherland,
                   Phys. Rev. B \textbf{12}, 3795 (1975).

\bibitem{Ori00} C. Itoi, S. Qin, and I. Affleck,
                   Phys. Rev. B \textbf{61}, 6747 (2000);
                E. Orignac, R. Citro, and N. Andrei,
                   Phys. Rev. B \textbf{61}, 11533 (2000).

\bibitem{Yam00} Y. Yamashita, N. Shibata, and K. Ueda,
                   J. Phys. Soc. Jpn. \textbf{69}, 242 (2000).

\bibitem{Li05}  Peng Li and Shun-Qing Shen,
                   Phys. Rev. B \textbf{72}, 214439 (2005).

\bibitem{Kol98} A. K. Kolezhuk and H.-J. Mikeska,
                   Phys. Rev. Lett. \textbf{80}, 2709 (1998).

\bibitem{Kol01} A. K. Kolezhuk, H.-J. Mikeska, and U. Schollw\"{o}ck,
                   Phys. Rev. B \textbf{63}, 064418 (2001).

\bibitem{Martins} M. J. Martins and B. Nienhuis,
                   Phys. Rev. Lett. \textbf{85}, 4956 (2000).

\bibitem{Mila}  F. Mila, B. Frischmuth, A. Deppeler, and M. Troyer,
                   Phys. Rev. Lett. \textbf{82}, 3697 (1999).

\bibitem{Kha01} G. Khaliullin, P. Horsch, and A. M. Ole\'s,
                   Phys. Rev. Lett. \textbf{86},   3879 (2001);
                   Phys. Rev. B     \textbf{70}, 195103 (2004).

\bibitem{Kug15} K. I. Kugel, D. I. Khomskii, A. O. Sboychakov,
                   and S.~V.~Streltsov,
                   Phys. Rev. B \textbf{91}, 155125 (2015).

\bibitem{You07} W.-L. You, Y.-W. Li, and S.-J. Gu,
                   Phys. Rev. E \textbf{76}, 022101 (2007).

\bibitem{Dag08} M. Daghofer, K. Wohlfeld, A. M. Ole\'s, E. Arrigoni,
                   and P. Horsch,
                   Phys. Rev. Lett. \textbf{100}, 066403 (2008);
                K.~Wohlfeld, M. Daghofer, A. M. Ole\'s, and P. Horsch,
                   Phys. Rev. B \textbf{78}, 214423 (2008).

\bibitem{Che09} G.-W. Chern and N. Perkins,
                   Phys. Rev. B   \textbf{80}, 220405(R) (2009);
                G.-W. Chern, N. Perkins, and G.~I.~Japaridze,
                   \textit{ibid.} \textbf{82}, 172408 (2010).

\bibitem{Dou05} B. Dou\c{c}ot, M. V. Feigel'man, L. B. Ioffe,
                   and A. S. Ioselevich,
                   Phys. Rev. B \textbf{71}, 024505 (2005);
                Z.~Nussinov and E. Fradkin,
                   \textit{ibid.} \textbf{71}, 195120 (2005).

\bibitem{Brz10} W. Brzezicki and A. M. Ole\'s,
                   Phys. Rev. B   \textbf{82}, 060401 (2010);
                   \textit{ibid.} \textbf{87}, 214421 (2013);
                   \textit{ibid.} \textbf{90}, 024433 (2014);
                L. Cincio, J. Dziarmaga, and A. M. Ole\'s,
                   \textit{ibid.} \textbf{82}, 104416 (2010).

\bibitem{Tro10} F. Trousselet, A. M. Ole\'s, and P. Horsch,
                   Europhys. Lett. \textbf{91},  40005 (2010);
                   Phys. Rev. B    \textbf{86}, 134412 (2012).

\bibitem{Kit06} A. Kitaev,
                   Ann. Phys. (NY) \textbf{321}, 2 (2006).

\bibitem{Nus15} Z. Nussinov and J. van den Brink,
                   Rev. Mod. Phys. \textbf{87}, 1 (2015).

\bibitem{Han}   T.-H. Han, J. S. Helton, S. Chu, D. G. Nocera,
                   J.~A.~Rodriguez-Rivera, C. Broholm, and Y. S. Lee,
                   Nature \textbf{492}, 406 (2012).

\bibitem{Maj69} C. K. Majumdar and D. Ghosh,
                   J. Math. Phys. \textbf{10}, 1388 (1969).

\bibitem{Yu96}  Y. Yu, G. M\"{u}ller, and V. S. Viswanath,
                   Phys. Rev. B \textbf{54}, 9242 (1996).

\bibitem{Ner11} A. A. Nersesyan and A. M. Tsvelik,
                   Phys. Rev. Lett. \textbf{78}, 3939 (1997);
                A. Nersesyan, G.-W. Chern, and N.~B.~Perkins,
                   Phys. Rev. B \textbf{83}, 205132 (2011).

\bibitem{Soos}  Z. G. Soos, S. Kuwajima, and J. E. Mihalick,
                   Phys. Rev. B \textbf{32}, 3124  (1985).

\bibitem{Spr86} G. Spronken, B. Fourcade, and Y. L\'{e}pine,
                   Phys. Rev. B \textbf{33}, 1886 (1986).

\bibitem{Gu07}  S.-J. Gu, G.-S. Tian, and H.-Q. Lin,
                   Chin. Phys. Lett.  \textbf{24}, 2737 (2007).

\bibitem{Brz12} W. Brzezicki, J. Dziarmaga, and A. M. Ole\'s,
                   Phys. Rev. Lett. \textbf{109}, 237201 (2012);
                   Phys. Rev. B \textbf{87}, 064407 (2013);
                   Acta Phys. Pol. A \textbf{126}, A-40 (2014).

\bibitem{Sir08} J. Sirker, A. Herzog, A. M. Ole\'s, and P.~Horsch,
                   Phys. Rev. Lett. \textbf{101}, 157204 (2008).

\bibitem{Plaq}  W. Brzezicki and A. M. Ole\'s,
                   Phys. Rev. B \textbf{90}, 024433 (2014).

\bibitem{Herzog} A. Herzog, P. Horsch, A. M. Ole\'s, and J. Sirker,
                   Phys. Rev. B \textbf{83}, 245130 (2011).

\bibitem{Wor63} M. Wortis,
                   Phys. Rev. \textbf{132}, 85 (1963).

\bibitem{Sch81} T. Schneider,
                   Phys. Rev. B \textbf{24}, 5327 (1981).

\bibitem{Brink} J. van den Brink, W. Stekelenburg, D. I. Khomskii,
                   G.~A.~Sawatzky, and K. I. Kugel,
                   Phys. Rev. B \textbf{58}, 10276 (1998).

\bibitem{Bal01} J. Ba{\l}a, A. M. Ole\'s, and G. A. Sawatzky,
                   Phys. Rev. B \textbf{63}, 134410 (2001).

\bibitem{Coj03} S. Cojocaru and A. Ceulemans,
                   Phys. Rev. B \textbf{67}, 224413 (2003).
                   


\end{thebibliography}
\end{document}